\documentstyle[12pt]{article}
\newcommand{\be}{\begin{eqnarray}}
\newcommand{\ee}{\end{eqnarray}}
\newcommand{\ba}{\begin{array}}
\newcommand{\ea}{\end{array}}
\newcommand{\bmat}[1]{\left(\!\begin{array}{#1}}
\newcommand{\emat}{\end{array}\!\right)}
\newcommand{\cM}{{\cal M}}
\newcommand{\cP}{{\cal P}}

\newcommand{\half}{{\textstyle{\frac{1}{2}}}}
\newcommand{\mint}[1]{\int\!\frac{d^3 #1}{(2\pi)^3}}
\newcommand{\Vpq}{V({\bf p} - {\bf q})}
\newcommand{\bfN}{{\bf N}}
\newcommand{\bfY}{{\bf Y}}
\newcommand{\tr}{{\rm tr}\,}
\newcommand{\Tr}{{\rm Tr}\,}
\newcommand{\bfp}{{\bf p}}
\newcommand{\bfq}{{\bf q}}
\newcommand{\bfph}{{\bf\hat{p}}}
\newcommand{\bfqh}{{\bf\hat{q}}}
\newcommand{\magp}{|{\bf p}|}
\newcommand{\magq}{|{\bf q}|}
\newcommand{\matel}[3]{\langle #1 | #2 | #3 \rangle}
\newcommand{\twocomp}[3]{{#1}^{\bigl(\!\!\!\! {\scriptstyle
\begin{array}{c} {\scriptstyle #2} \\[-1.2ex] {\scriptstyle #3}
\end{array}} \!\!\!\!\bigr) }}
\newcommand{\olin}[1]{\overline{#1}}
\newcommand{\wtil}[1]{\widetilde{#1}}
\newcommand{\Psiz}{\stackrel{0}{\Psi}}
\newcommand{\Lambdaz}{\stackrel{0}{\Lambda}}
\newcommand{\cf}{{\em cf.}}
\newcommand{\ie}{{\em i.e.}}
\newcommand{\etal}{{\em et al.}}
\begin{document}
%
%
\thispagestyle{empty}
\renewcommand{\thefootnote}{\alph{footnote}}
\rightline{UNITUE-THEP-17/93}
\rightline{hep-ph/9402236}
\vspace{1.5cm}
\begin{center}
{\large\bf HEAVY--LIGHT MESONS IN A BILOCAL EFFECTIVE \\[.5ex] THEORY}
\vspace{1.5cm} \\
{\large Yu. L. Kalinovsky}\footnote[1]{Permanent address: Bogoliubov
Laboratory of Theoretical Physics, Joint Institute for Nuclear Research,
Dubna, Head Post Office, P.O. Box 79, 101000 Moscow,
Russia}\footnote[2]{e-mail address: yura@darss.mpg.uni-rostock.de}
\vspace{0.4cm} \\
{\em MPG AG ``Theoretische Vielteilchenphysik'' \\ Universit\"at Rostock \\
Universit\"atsplatz 1, D-18051 Rostock, Germany}
\vspace{0.7cm} \\
{\large C. Weiss}\footnote[3]{Supported by a research fellowship of the
Deutsche Forschungsgemeinschaft (DFG)}\footnote[4]{e-mail address:
weiss@mailserv.zdv.uni-tuebingen.de}
\vspace{0.4cm} \\
{\em Institut f.\ Theoretische Physik \\ Universit\"at T\"ubingen \\
Auf der Morgenstelle 14, D-72076 T\"ubingen, Germany}
\end{center}
\vspace{1cm}
\begin{abstract}
\noindent
Heavy--light mesons are described in an effective quark theory with a
two--body vector--type interaction. The bilocal interaction is taken to be
instantaneous in the rest frame of the bound state, but formulated
covariantly through the use of a boost vector. The chiral symmetry of the
light flavor is broken spontaneously at mean field level. The framework for
our discussion of bound states is the effective bilocal meson action
obtained by bosonization of the quark theory. Mesons are described by
3--dimensional wave functions satisfying Salpeter equations, which exhibit
both Goldstone solutions in the chiral limit and heavy--quark symmetry for
$m_Q\rightarrow\infty$. We present numerical solutions for pseudoscalar
$D$-- and $B$--mesons. Heavy--light meson spectra and decay constants are
seen to be sensitive to the description of chiral symmetry breaking
(dynamically generated vs.\ constant quark mass).
\end{abstract}
\vfill
\renewcommand{\thefootnote}{\arabic{footnote}}
\eject
\section{Introduction}
One of the most interesting problems of hadronic physics is the study of
hadrons consisting of heavy and light quarks. For example, $B$--meson
decays play an important role in determining the elements of the
Kobayashi--Maskawa matrix, including the $CP$--violating phase
\cite{ali_91}. Furthermore, rare decays of heavy mesons may indicate
deviations from the standard model. For the description of physical
(hadronic) decay processes one needs to know the meson wave functions and
form factors.
\par
Heavy--light systems are also challenging from the point of view of strong
interactions. The physics of light flavors is largely governed by the
chiral symmetry of QCD and its spontaneous breaking in the vacuum. In
particular, this entails the existence of light pseudoscalar Goldstone
bosons, the pions. A simple model exemplifying the dynamical mechanism of
chiral symmetry breaking is the Nambu--Jona-Lasinio model, which is based
on a local two--body interaction between quarks
\cite{volkov_83,reinhardt_86,klimt_90}. For heavy flavors $(c, b)$ the
situation is quite different. In this realm chiral symmetry breaking plays
a minor role. The strong dynamics of heavy quarks simplifies because of the
fact that they behave essentially like classical particles, \ie , their
off--shellness in a bound state is small compared to the quark mass. Thus,
heavy quarkonia ($c\bar c, b\bar b $) are successfully described by
non-relativistic potential models using a Coulomb--type interaction at
short distances and a linear confinement potential at large distances
\cite{eichten_78,mukherjee_93}. Heavy--light hadrons occupy an intermediate
position. In heavy--light bound states, the velocity of the heavy quark is
conserved in the limit $m_Q\rightarrow\infty$, and the mass spectrum
becomes independent of the heavy--quark spin. Recently, a general effective
theory for heavy--light mesons has been constructed on the basis of both
the chiral symmetry of the light flavors and the heavy--quark limit for the
heavy flavors \cite{wise_91,eichten_90}. However, the coefficients of this
effective Lagrangian are not determined by the symmetries, but by the
details of the underlying dynamics, QCD. Thus, for a quantitative
understanding of heavy--light systems it is necessary to consider a
suitable approximation to QCD at quark level. This approximation must take
into account certain qualitative features of QCD, most importantly quark
confinement and the spontaneous breaking of chiral symmetry. Moreover, it
should incorporate the Coulomb interaction between quarks at short
distances. These requirements make it necessary to consider models with
non-local effective interactions, which can simulate both short--distance
and non-perturbative long--distance effects.
\par
Here we consider the possibility of describing heavy--light mesons in the
framework of a bilocal effective quark theory \cite{kleinert_76,ebert_78}.
To incorporate the spontaneous breaking of chiral symmetry, we use a
Lorentz--vector effective interaction motivated by QCD. A special feature
of our approach is that we take the interaction to be instantaneous in the
rest frame of the meson bound state \cite{kalin1_89,perv_90,kalin_90}.
There are a number of reasons for this choice. Such an interaction leads to
a transparent description of mesons in terms of Salpeter wave functions,
which satisfy simple Schr\"odinger--type equations. Nevertheless, the
relativistic kinematics as well as effects of spontaneous chiral symmetry
breaking are incorporated exactly in the form of Foldy--Wouthuysen factors
arising in the reduction of the Bethe--Salpeter equation. Such a model
reduces in the non-relativistic limit to the potential model for heavy
quarkonia ($Q\bar Q$). Furthermore, instantaneous interactions in the form
of the Coulomb gauge have long been used to study chiral symmetry breaking
in a quasiparticle picture of the QCD vacuum
\cite{finger_82,leyaouanc_84,adler_84,alkofer_88}.
\par
In general, potential models do not possess full relativistic invariance
because the Fock space is restricted to $q \bar q$--pairs and an
instantaneous interaction is assumed. Nevertheless, in our model the
interaction is written in a relativistically covariant form through the use
of a boost vector proportional to the bound--state total momentum.
Physically, this corresponds to the intuitive picture of a potential moving
together with the bound state, as is used to describe moving atoms in QED.
Moreover, for heavy--light bound states this boost vector coincides with
the 4--velocity of the heavy quark, which provides a natural connection of
this approach to heavy--quark effective theory. This relativistic
formulation allows us to define Bethe--Salpeter amplitudes and wave
functions for moving particles, which is crucial for the calculation of
formfactors or decay matrix elements leading to the Isgur--Wise functions.
A description of heavy meson and baryon weak decays in the heavy quark
limit based on Bethe--Salpeter wave functions has been developed by Hussain
\etal\ \cite{hussain_90}. Recently, Dai \etal\ have considered a covariant
instantaneous interaction in this context \cite{dai_93}. In contrast to
these approaches we do not take the heavy--quark limit from the start but
formulate the description of bound states for arbitrary quark masses,
discussing the simplifications arising in the heavy quark limit afterwards.
An advantage of this covariant approach is precisely the possibility to
describe in a unified manner both light and heavy mesons, using appropriate
potentials. For example, for light quarks one may employ a separable
approximation to an intermediate--range potential in momentum space. In
this case one obtains a consistently regularized version of the
Nambu--Jona--Lasinio model, in which the 3--dimensional momentum space
cutoff moves along with the bound state \cite{kalin2_89}. We should point
out that the relativistic formulation using the boost vector is useful also
in the light quark sector, {\em e.g.}\ in defining the pion decay constant
\cite{leyaouanc_85}.
\par
Nowak \etal\ have established the general relation between dynamics at
quark level and heavy quark effective theory by performing a gradient
expansion of the fermion determinant in the heavy quark limit
\cite{nowak_93}. The bilocal meson action derived in our approach may serve
as an explicit realization of their ideas. However, we shall consider here
the full momentum--dependent theory and do not restrict the effective
action to the long--wavelength limit.
\par
In this paper, we first develop the formal framework for the description of
meson bound states in a bilocal effective model with a covariant
instantaneous interaction, which takes into account the spontaneous
breaking of chiral symmetry. We then apply this model to the study of
light--light and heavy--light pseudoscalar mesons. Specifically, we want to
demonstrate that the momentum dependence of the constituent quark mass,
which is a consequence of the dynamical nature of chiral symmetry breaking,
has important effects on the spectrum and decay constants of heavy--light
mesons. The description of meson transitions and form factors in this
unified approach will be given in the future \cite{weiss_94}.
\par
This paper is organized as follows. In sect.\ 1 we formulate the effective
quark theory and introduce the instantaneous interaction in the covariant
formulation \cite{kalin1_89,perv_90}. We discuss the phenomenological
3--dimensional potentials considered in the applications. We then bosonize
the model and obtain an effective bilocal meson theory. The
Schwinger--Dyson equation for the vacuum and the Bethe--Salpeter equation
for the meson fluctuations are derived from the effective meson action. We
outline the calculation of matrix elements between bound states needed to
describe meson decay constants or semileptonic decays. In these derivations
we assume a general covariant interaction kernel. In sect.\ 2 we then
discuss in detail the equations describing bound states for the case of a
covariant instantaneous interaction. The Schwinger--Dyson equation
describes the quark quasiparticle spectrum. The Bethe--Salpeter equation
for mesons is rewritten in terms of a Salpeter wave function. We make use
of a Foldy--Wouthuysen transformation to simplify the form of the resulting
equations. In sect.\ 3, we then apply the model to the description of
light--light and heavy--light mesons. We discuss the chiral limit, $m_q
\rightarrow 0$, and study the behavior of the pseudoscalar meson in this
regime. We then investigate heavy--light bound states ($Q\bar q$). In the
heavy quark limit, $m_Q \rightarrow\infty$, reduced bound state equations
are obtained, which exhibit the heavy--quark spin symmetry. We present
numerical solutions of the full equations for pseudoscalar $D$-- and
$B$--mesons. Specifically, we compare meson properties calculated with the
momentum--dependent quark mass from the Schwinger--Dyson--equation with
those obtained using a constant light quark mass. We find that heavy--light
meson properties are sensitive to whether chiral symmetry breaking for the
light flavor is described self--consistently or by a constant constituent
quark mass. A summary and an outlook are given in sect.\ 4.
\par
Appendix A.1 deals with the partial--wave analysis of the meson Salpeter
equation. In appendix A.2 we describe the numerical solution of the
partial--wave equations using the Multhopp method, which has been very
sucessful in the context of constituent quark models \cite{boukraa_89}. In
particular, we consider it necessary to comment on the complications
presented by the momentum--dependent Foldy--Wouthuysen factors in the bound
state equations of our model. Appendix B outlines the evaluation of the
matrix elements for the normalization of meson wave functions and the
calculation of meson decay constants from the bilocal effective meson
action.
\section{Effective quark theory with instantaneous interaction}
The basis of our description is a quark theory with an effective two--body
interaction in the color--octet channel. It is defined by the action of the
quark field,
\be
W &=& \int\!  d^4 x\; \bar{q}(x) G_0^{-1} (x) q(x) \; -\;
\half g^{2}  \int\!\int\! d^4 x\, d^4 y\; j^{\mu a}(x)
\hat{D}^{ab}_{\mu\nu}(x - y) j^{\nu b}(y) .
\label{action_q}
\ee
Here, $G_0$ is the free quark Green function,
\be
G_0^{-1} &=& i \rlap/\partial - \hat{m}^0 ,
\ee
where $\hat{m}^0 = \mbox{diag}(m^0_1 , \ldots , m^0_{N_f})$ is the quark
current mass matrix, with $N_f$ the number of flavors. The quark color
current is given as
\be
j^a_\mu (x) &=& \bar{q} (x)
\biggl( \frac{\lambda^a}{2} \biggr) \gamma_\mu q (x) \,\,,
\ee
where $\lambda^a$ are the Gell--Mann matrices of $SU(3)_{\rm c}$.
The bilocal interaction kernel,
\be
\hat{D}^{ab}_{\mu\nu}(x-y) &\equiv& \delta^{ab} g_{\mu\nu} D(x-y) ,
\ee
can be thought of as an effective gluon propagator, which describes part of
the non-abelian effects of QCD in a phenomenological way. Note that the
interaction is of Lorentz--vector type, as motivated by QCD, and thus
chirally invariant.
\par
For the purpose of describing meson bound states we rewrite the interaction
term of eq.(\ref{action_q}) in the form
\be
\int\!\int\! d^4 x\, d^4 y\;  q_{B}(y) \bar{q}_{A}(x) \hat{K}_{AB,CD}(x - y)
q_{D}(x) \bar{q}_{C}(y) ,
\ee
with the kernel
\be
\hat{K}_{AB, CD}(x - y) &=& {\gamma}_{ru}^{\mu}(\gamma_{\mu})_{ts}
\,\sum_{a=1}^{8} {\frac {\lambda_{\alpha \delta}^{a}} {2}}
{\frac {\lambda_{\gamma \beta} ^{a}} {2}}\,
\delta_{il} \delta_{kj} \; \frac{g^2}{2} D(x - y) .
\label{kernel_K}
\ee
Here, $A = \{r, \alpha, i\}, \ldots , D = \{ u, \delta , l \}$ are a
short--hand notation for the Dirac spinor, color and and flavor indices. In
the following we want to study meson $(q \bar q )$ bound states. We
therefore make the well--known color Fierz rearrangement \cite{cahill_89}
\be
\sum_{a=1}^{8}
\lambda^a_{\alpha\delta}
\lambda^a_{\gamma\beta} &=& \frac{4}{3}
\delta_{\alpha \beta} \delta_{\gamma \delta}
\; +\; \frac{2}{3} \sum_{\rho=1}^{3}
\epsilon_{\rho \alpha \gamma}  \epsilon^{\rho \beta \delta } ,
\ee
where $\epsilon_{\alpha\beta\gamma}$ is the antisymmetric Levi--Civita
tensor. This identity allows to rewrite the interaction completely into the
attractive color--singlet ($ q \bar{q} $) and antitriplet ($q q$) channels,
while the repulsive color--octet and sextet channels are absent in a
natural way. We consider only the color--singlet part of
eq.(\ref{kernel_K}). The relevant part of the action, eq.(\ref{action_q}),
can then be represented in the form
\be
W &=& \int\!\int\! d^4 x\, d^4 y\;
\biggl\{
\bigl( q(y) \bar{q}(x) \bigr) \,
\bigl( - G_0^{-1}(x) \bigr) \delta (x-y)
\label{W_color_singlet} \\
&& \hspace{2em} + \;\frac{1}{2N_{c}}
\Bigl[
\bigl( q(y) \bar{q}(x) \bigr)\, K (x - y)\, \bigl( q(x) \bar{q}(y) \bigr)
\Bigr]
\biggr\} ,
\nonumber
\ee
with the color--singlet interaction kernel
\be
K (x - y) &=& \gamma^\mu \otimes \gamma_\mu\, g^2 D(x - y) .
\label{K_color_singlet}
\ee
Here, the bilinear $\bigl( q(y) \bar{q}(x) \bigr)$ is contracted over color
indices, and $N_{\rm c} = 3$ is the number of colors.
\par
If the model action eq.(\ref{action_q}) is considered as a Euclidean field
theory with a covariant gluon propagator, eq.(\ref{action_q}) is known as
the so-called Global Color Model \cite{praschifka_87}. In that approach,
bound states are studied through the Euclidean correlation functions.
Similar covariant models have been investigated in
\cite{munczek_92,smekal_91}. Here, we consider eq.(\ref{action_q}) in a
different context. For reasons already stated above we wish to have an
effective theory with an instantaneous interaction. Such an interaction
leads to a simple description of bound states in terms of 3--dimensional
wave functions in Minkowski space, which for heavy-heavy bound states
naturally reduces to the successful non-relativistic potential model. The
concept of an instantaneous interaction can be formulated covariantly by
letting the potential move along with the bound state.
\par
Consider a bound state of two quarks interacting through a bilocal
effective interaction of the form eq.(\ref{K_color_singlet}). From the
principle of translational invariance it follows that one can separate the
center--of--mass motion of the bound state from the relative motion in the
form of a plane wave. The momentum of the c.o.m.\ motion equals the total
momentum of the quark pair, $\cP_{\mu}$. Given this, one can more or less
arbitrarily define a conjugate coordinate, $X_{\mu}$, representing the
absolute position in space--time and a relative coordinate, $z_\mu$, to
describe the internal structure of the bound state. The condition is that
under a translation by a constant vector $a$, $x\rightarrow x + a, y
\rightarrow y + a$, these coordinates transform as $X\rightarrow X + a, z
\rightarrow z$. This is satisfied for any linear combination $X = \alpha x
+ (1- \alpha )y$ and $z = x - y$. We shall take $\alpha = \half$ in the
following\footnote{The definition of the center--of--mass coordinate as $X
= (m_1 x + m_2 y)/(m_1 + m_2)$ has significance only in the
non-relativistic limit. In a relativistic theory it is in general
impossible to write the two--body hamiltonian as the sum of two terms
describing c.o.m.\ and relative motion.}.
\par
We now want to substitute the general bilocal interaction kernel of
eq.(\ref{K_color_singlet}) by an instantaneous one, \ie , by a potential.
This can be done in a relativistically covariant way by replacing
eq.(\ref{K_color_singlet}) by
\be
K^{\eta}(x,y) &=& K^{\eta}\Biggl( x - y \;\rule[-2ex]{.2mm}{5ex}\;
\frac{x+y}{2} \Biggr)
\; =\; - \rlap/\eta \otimes \rlap/\eta\, V( z^{\perp} )\,
\delta ( z^{\vert\vert} ) .
\label{K_eta}
\ee
Here,
\be
z_{\mu}^{\vert\vert} &=& \eta_{\mu} ( z\!\cdot\!\eta ) , \hspace{2em}
z_{\mu}^{\perp} \; = \; z_{\mu} - z_{\mu}^{\vert\vert} ,
\label{z_perp}
\ee
and $\eta_{\mu}$ is a boost vector proportional to the total momentum
eigenvector of the bound state, $\cP_{\mu}$,
\be
\eta_{\mu} &=& \frac{\cP_{\mu} }{\sqrt{\cP^2}} , \hspace{2em}
\eta^2 \; =\; 1 , \hspace{2em}
\rlap/\eta \; = \; \eta_{\mu} \gamma_{\mu}.
\ee
In eq.(\ref{K_eta}), the $\delta$--function, $\delta (z\!\cdot\!\eta )$,
guarantees the instantaneousness of the exchange interaction in the rest
frame of the bound state. The transversality of the exchange interaction is
ensured by the fact that the 3--dimensional potential is a function only of
the perpendicular part of the relative coordinate, $V( z^{\perp})$. The
sign in eq.(\ref{K_eta}) has been chosen such that an attractive
3--dimensional potential leads to to an attractive interaction in the
color--singlet (and antitriplet) channel. Eq.(\ref{K_eta}) describes a
potential moving together with the bound state. This form of interaction
leads to bound--state amplitudes with well--defined Lorentz transformation
properties. We remark that one can arrive at the form eq.(\ref{K_eta}) by
studying moving bound states within a canonical quantization approach to
gauge theories \cite{perv_85}.
\par
For the bound state at rest one has $\vec\cP = 0$, so that $\eta_{\mu} =
(1, \vec{0})$. In this frame the kernel takes the form
\be
K^{(1, \vec{0})} (x, y) &=& - \gamma_0 \otimes \gamma_0
V({\bf x} - {\bf y}) \delta (x_0 - y_0 ) .
\label{K_rest_frame}
\ee
In electrodynamics this kernel corresponds to the usual Coulomb gauge, with
$V({\bf x} - {\bf y})$ the Coulomb potential. The form
eq.(\ref{K_rest_frame}) has been widely used as a model to describe the
breaking of chiral symmetry in strong interactions
\cite{finger_82,leyaouanc_84,adler_84,alkofer_88}.
\par
Let us now assume that the bound state contains a heavy quark ($c, b$) and
a light antiquark ($\bar{u}, \bar{d}, \bar{s}$). In the limit of heavy
quark effective theory, $m_Q \rightarrow \infty$, one has
\be
\eta_{\mu} &=& \frac{\cP_{\mu}}{\sqrt{\cP^{2}}}\; \rightarrow \; v_{\mu} ,
\ee
where $v$ is the 4--velocity of the heavy quark, $v^{2} = 1$. Thus, in this
limit the interaction kernel, eq.(\ref{K_eta}), takes the form
\be
K^{v}(x - y) &=& - \rlap/{v} \otimes \rlap/{v}
V(z^\perp ) \delta (v\!\cdot\!z) ,
\ee
where $z^\perp = z - v (v\!\cdot\!z)$. The fact that the interaction kernel
explicitly involves the heavy quark 4--velocity leads to a natural relation
of this potential approach with heavy--quark effective theory
\cite{wise_91,eichten_90}.
\par
The transverse potential, $V(z^\perp )$, is a phenomenological input to
this model. Its form may be chosen depending on the type of bound state one
wishes to study (light--light, heavy--light). We shall in the following
work with a potential, which in the rest frame is of the form
\be
V({\bf x} - {\bf y}) &=&
- \frac{4}{3} \frac{\alpha_s}{r} \, +\, \sigma^2 r ,
\label{potential}
\ee
with $r = |{\bf x} - {\bf y}|$. Note that when expressing the interaction
in an arbitrary frame one must take into account that $z^\perp$ is subject
to a non-Euclidean metric. The first term of eq.(\ref{potential}) is the
Coulomb potential describing one--gluon exchange. In the investigations in
this paper we shall take $\alpha_s$ as constant. One could also employ in
this approach an asymptotically free potential with $\alpha_s = \alpha_s
(\magq^2 )$ \cite{adler_84,alkofer_88}. The linear potential in
eq.(\ref{potential}) implements quark confinement in a phenomenological
way. In our study of heavy--light mesons we use parameters determined
within the non-relativistic potential model for charmonium, $\alpha_s \sim
0.3\ldots 0.5, \, \sigma \sim 0.4\,{\rm GeV}$
\cite{eichten_78,mukherjee_93}. For reference, we shall also consider the
oscillator model of Le Yaouanc \etal , $V({\bf x} - {\bf y}) = V_0^3 r^2$,
with $V_0 \sim 0.2\ldots 0.4\,{\rm GeV}$, which gives a good overall fit to
the light meson mass spectrum \cite{leyaouanc_85}. We remark that one may
include in eq.(\ref{potential}) also an intermediate--range attractive
potential, which can be thought of as a crude representation of instanton
effects \cite{hirata_89,muenz_93}. Another possibility is a separable
potential in momentum space, which leads to a Nambu--Jona--Lasinio--type
model. Such an interaction has been considered within this covariant
approach in \cite{kalin2_89}.
\section{Bound states}
\subsection{Hadronization}
For the discussion of meson bound states it is convenient to ``hadronize''
the quark theory, \ie , to formally rewrite it as an effective meson theory
\cite{reinhardt_86,ebert_78}. For the general discussion here we assume for
simplicity a general covariant interaction kernel. The equations obtained
will then be specified to the case of a co-moving instantaneous
interaction, eq.(\ref{K_eta}), in the following two sections.
\par
Let us consider the functional integral
\be
Z &=& \int {\cal D}q {\cal D}\bar{q} \; \exp i W[\bar q, q] .
\ee
Here, $W$ is the color--singlet part of the effective quark action,
eq.(\ref{W_color_singlet}), which in symbolic notation can be written as
\be
W[\bar q, q] &=&
\bigl( q \bar{q}, - G_0^{-1}  \bigr)\; +\; \frac{1}{2N_{c}}
\bigl(q \bar{q}, K \, q \bar{q} \bigr) .
\ee
After integrating over the quark fields with the help of the Legendre
transform one obtains
\be
Z &=& \int {\cal D}\cM \; \exp i W_{\rm eff} [ \cM ] ,
\ee
with the effective meson action
\be
W_{\rm eff} [ \cM ] &=&
N_c
\biggl\{
- \half ( \cM, K^{-1} \cM )
-i \Tr \log ( -G_0^{-1} + \cM )
\biggr\} .
\label{W_eff}
\ee
Here, ${\cal M} = {\cal M}_{ij} (x, y) \sim q_i (x)\bar q_j (y)$ is a
bilocal meson (color--singlet) field. It has the structure of a matrix in
Dirac spinor and flavor space. In eq.(\ref{W_eff}), the symbol $\Tr$
implies integration over the continuous variables as well as the traces
over spinor and flavor indices.
\par
The vacuum of the effective meson theory is given as the minimum of the
effective action, eq.(\ref{W_eff}). The condition of minimum reads
\be
K^{-1}\cM +i \frac{1}{-G_0^{-1}+\cM } &=& 0 .
\label{W_eff_minimum}
\ee
The vacuum solution of this equation is translationally invariant. Let us
denote it by ($\Sigma - \hat{m}^0 $). Then we obtain from
eq.(\ref{W_eff_minimum}) the Schwinger--Dyson equation
\be
\Sigma &=& \hat{m}^0 + i K G_{\Sigma} ,
\label{sde}
\ee
where
\be
G^{-1}_{\Sigma} &=& i \rlap/\partial - \Sigma .
\ee
Mesons are described as fluctuations of ${\cal M}$ around the vacuum
configuration. Expanding the action, eq.(\ref{W_eff}), around the minimum,
with $ \cM = (\Sigma - \hat{m}^0 ) + \cM^{\prime}$, one obtains
\be
W_{\rm eff} [\cM ] &=& W_{\rm eff} [ \Sigma ] \nonumber \\
&+& N_{c} \biggl\{
- \frac{1}{2} ( \cM^{\prime}, K^{-1}
\cM^{\prime} ) - \frac{i}{2}
\Tr ( G_{\Sigma} \cM^{\prime} )^2
- i \sum^{\infty}_{n=3} \frac{1}{n} \Tr
(- G_{\Sigma} \cM^{\prime} )^n
\biggr\} .
\ee
The vanishing of the second variation of this effective action with respect
to $ \cM^{\prime}$,
\be
\frac{\delta^2 W_{\rm eff}}{ \delta \cM^{\prime}  \delta \cM^{\prime} }
{\Bigg \vert}_{ \cM^{\prime} =0 } \cdot \Gamma = 0 \,\, , \nonumber
\ee
leads to the homogeneous Bethe--Salpeter equation for the vertex function,
$\Gamma$,
\be
\Gamma
= -i K (G_{\Sigma}  \Gamma  G_{\Sigma}) .
\label{bse}
\ee
This equation describes the bound state spectrum. It corresponds to the
usual Bethe--Salpeter equation in ladder approximation.
\par
Given the solutions of eqs.(\ref{sde}, \ref{bse}) describing the bound
state spectrum, one can calculate matrix elements between on-shell bound
states describing meson decays, semi\-leptonic processes, {\em etc.}, in
the framework of the effective meson action, eq.(\ref{W_eff}). The
bosonized action summarizes these processes in a concise way and greatly
facilitates the evaluation of the corresponding matrix elements. One
formally expands the bilocal field, $\cM$, in bound state amplitudes,
\be
\lefteqn{
\cM (x,y) \; = \; \cM \biggl( x - y \;\rule[-2ex]{.2mm}{5ex}\;
\frac{x+y}{2} \biggr)
\; = \; \sum_{H} \int \frac{d\cP^3}{(2 \pi)^{3/2} \sqrt{2 \omega_{H}}}}
\nonumber \\
&& \times \int \frac{d^{4}q}{(2 \pi)^{4}} e^{iq(x-y)} \Bigl\{ e^{i \cP
\frac{x+y}{2}} a^{+}_{H}(\vec{\cP})
\Gamma_{H}(q \vert \cP) + e^{-i \cP \frac{x+y}{2}} a_{H}(\vec{\cP})
\olin{\Gamma}_{H}(q \vert \cP) \Bigr\} .
\label{M_expansion}
\ee
The sum runs over the set of quantum numbers, $H$, of hadrons contributing
to the bilocal field. Here, the individual bound states have mass $M_{H}$
and total 4--momentum $\cP = (\omega_{H}, \vec{\cP} )$, with
$\omega_{H}(\vec{\cP}) = (\vec{\cP}^{2} + M_{H}^{2})^{1/2}$. The amplitudes
$\Gamma_H (q|\cP )$ and $\olin{\Gamma}_H (q|\cP ) = \Gamma_H (q| -\cP )$
are on--shell solutions of the Bethe--Salpeter equation, eq.(\ref{bse}),
with 3--momentum $\vec{\cP}$. The coefficients $a_H^+ (\vec{\cP}), a_H
(\vec{\cP})$ may then be interpreted as creation and annihilation operators
of physical (on--shell) mesons, and matrix elements can be calculated as
usual. In particular, the bound states amplitudes are normalized by the
requirement that the matrix element of the quadratic (free) part of the
effective action, eq.(\ref{W_eff}),
\be
W_{\rm eff}^{(2)} &=& -i \half N_{c} \Tr (G_{\Sigma} \cM )^{2}
\label{W_2}
\ee
have the normalization corresponding to a physical (elementary) particle.
This is simply the statement of the correct relativistic dispersion law of
the c.o.m.\ motion of the bound state\footnote{For a covariant
instantaneous interaction, eq.(\ref{K_eta}), the interaction does not
contribute to the normalization of the bound state, \cf\ appendix B.}.
\par
In order to describe meson weak decays, we couple the theory
eq.(\ref{action_q}) to an external weak leptonic current. The corresponding
lagrangian at quark level is given by
\be
{\cal{L}}_{\rm semi} &=& \frac{G_{F}}{\sqrt{2}}
\{V_{ij}(\bar{Q}_{i}(x) O_{\mu} q_{j}(x)) l_{\mu}(x)  \; + \; {\rm h.c.} \} .
\label{weak_interaction}
\ee
Here, $l_\mu$ is the leptonic current,
\be
l_{\mu}(x) &\equiv& \bar{l}(x) O_{\mu}\nu_{l}(x), \hspace{2em} l\; =\;
(e, \mu, \tau ), \hspace{2em} \nu_l\; =\; (\nu_e , \nu_\mu, \nu_\tau ) ,
\\
O_{\mu} &=& \gamma_{\mu}(1+\gamma_{5}) , \nonumber
\ee
$V_{ij}$ are the elements of the Kobayashi--Maskawa matrix, and $G_F$ is
the Fermi constant. In eq.(\ref{weak_interaction}), $Q$ denotes the column
of $(u,c,t)$--quarks, $q$ the $(d,s,b)$--quarks. On the level of the
effective meson action, eq.(\ref{W_eff}), the electroweak coupling can be
incorporated by shifting the bilocal field by the local leptonic current,
\be
\cM_{ij} (x,y) &\rightarrow& \cM_{ij} (x,y) \, +\, \hat{L}_{ij}(x,y), \\
\hat{L}_{ij}(x,y) &=& \frac{G_{F}}{\sqrt{2}} \delta^{4}(x-y) V _{ij}
O^{\mu} l_{\mu}(x) e^{i\cP_{L} \frac{x+y}{2}} ,
\ee
where $\cP_{L}$ is the momentum of the leptonic pair. After this shift, the
part of eq.(\ref{W_2}) describing meson decay into leptons is
\be
W_{\rm semi}^{(2)} &=& -i N_{c} \Tr (G_\Sigma \cM G_\Sigma \hat{L} ) .
\label{W_semi_2}
\ee
Semileptonic decays of mesons, which for heavy--light mesons are
parametrized in terms of the Isgur--Wise functions \cite{isgur_89}, are
mediated by
\be
W_{\rm semi}^{(3)} &=& i N_{c} \Tr
(G_\Sigma \cM G_\Sigma \cM G_\Sigma \hat{L} ) .
\label{W_semi_3}
\ee
\par
In the following we consider in detail the Schwinger--Dyson--equation,
eqs.(\ref{sde}), and the Bethe--Salpeter--equation, eq.(\ref{bse}), for a
covariant instantaneous interaction, eq.(\ref{K_eta}). In particular, bound
states will be described by 3--dimensional wave functions. The evaluation
of meson observables then proceeds in general as follows. Given the
solution for the bound state wave function in the rest frame, one can
reconstruct the Bethe--Salpeter amplitude in the rest frame,
\cf\ eqs.(\ref{Psi_undressed}, \ref{Pi_0}, \ref{schroedinger_eq}) below. By
virtue of the relativistic formulation of the interaction,
eq.(\ref{K_eta}), the bound state amplitude transforms covariantly and may
be boosted to an arbitrary velocity. Matrix elements of the vertices
eqs.(\ref{W_semi_2}, \ref{W_semi_3}) {\em etc.}\ between moving bound
states can then be calculated from the expansion of the bilocal field,
eq.(\ref{M_expansion}). The evaluation of the matrix element for
pseudoscalar meson decay in this approach is discussed in appendix B.
\subsection{The quark spectrum within a meson}
For an instantaneous interaction of the form eq.(\ref{K_eta}) the
Schwinger--Dyson equation, eq.(\ref{sde}), describes the spectrum of quark
quasiparticles in the rest frame of the meson. In momentum space it
takes the form
\be
\Sigma(p^\perp ) &=& \hat{m}^0 - i \int \frac{d^{4}q}{(2 \pi)^4}
V(p^\perp - q^\perp )
\rlap/\eta G_{\Sigma} (q) \rlap/\eta .
\label{sde_eta}
\ee
Here, the 4--dimensional Fourier transform of the interaction kernel,
eq.(\ref{K_eta}), depends only on the transverse part of the relative
momenta,
\be
V(p^\perp - q^\perp ) &=& \int d^{4}x\, e^{-i (p-q)x} V(x^{\perp})
\delta (x\!\cdot\!\eta) .
\label{V_fourier}
\ee
In the following we consider eq.(\ref{sde_eta}) in the rest frame, where
$\eta_{\mu} = (1, \vec{0}),\, \rlap/\eta = \gamma^0$, and
eq.(\ref{V_fourier}) coincides with the usual 3--dimensional Fourier
transform. Assuming that $ \Sigma({\bf p}) $ is diagonal in flavor,
$\Sigma = {\rm diag}\, (\Sigma_{1}, \ldots \Sigma_{N_{f}})$,
eq.(\ref{sde_eta}) splits into identical equations for the $\Sigma_n$ with
bare mass $m^0_n, \, n = 1,\ldots N_{f}$. We shall omit the flavor index
on $\Sigma$ in the following. The quark self--energy in the rest frame has
a scalar and a vector part,
\be
\Sigma({\bf p}) &=& A(\magp )\,\magp \,
+\, B(\magp )\,\bfp\!\cdot\!\vec{\gamma} .
\label{Sigma_A_B}
\ee
In the following it will be convenient to make a polar decomposition of the
quark energy in the form
\be
\Sigma ({\bf p}) \, +\, \bfp\!\cdot\!\vec{\gamma} &=& E(\magp )\,
S^2 ({\bf p}) ,
\ee
where $S^2 (\bfp )$ is the square of a Foldy--Wouthuysen matrix,
\be
S^{\pm 2} (\bfp ) &=& \sin\phi (\magp ) \,\pm\, \bfph\!\cdot\!\vec{\gamma}
\cos\phi (\magp ),
\label{S2} \\[.5ex]
E(\magp ) \sin\phi (\magp ) &=& A(\magp )\,\magp , \hspace{2em}
E(\magp ) \cos\phi (\magp ) \; =\; (1 + B(\magp ))\,\magp .
\label{A_B}
\ee
Here, $0 \leq \phi (\magp ) \leq \half\pi$ is the so--called chiral angle.
(In the quasiparticle language this angle defines the rotation of the
massive quasiparticle spinors relative to the free quark spinors in the
vacuum of broken chiral symmetry \cite{finger_82,leyaouanc_84}.) In this
parametrization the quark propagator, $G_{\Sigma}(q)$, becomes
\be
G_{\Sigma}(q) \; =\;
\frac{1}{\rlap/{q} - \Sigma({\bf q})}  &=&
\biggl(
\frac{\Lambda_+({\bf q})}{ q_0 - E({\bf q}) +i \epsilon } +
\frac{\Lambda_-({\bf q})}{ q_0 + E({\bf q}) -i \epsilon }
\biggr) \gamma_0
\label{G_Sigma} \\
&=& \gamma_0 \biggl(
\frac{\olin{\Lambda}_+({\bf q})}{ q_0 -E({\bf q}) +i \epsilon } +
\frac{\olin{\Lambda}_-({\bf q})}{ q_0 +E({\bf q}) -i \epsilon }
\biggr) . \nonumber
\ee
Here, the matrices $\Lambda_\pm (\bfq),\, \olin{\Lambda}_\pm (\bfq)$ are
defined as
\be
\Lambda_{\pm}({\bf q}) &=&
S^{-1}({\bf q}) \Lambdaz_{\pm} S({\bf q}) ,
\hspace{2em}
\olin{\Lambda}_{\pm}({\bf q}) \; =\;
S({\bf q}) \Lambdaz_{\pm} S^{-1}({\bf q}) ,
\label{projectors}
\ee
where
\be
\Lambdaz_{\pm} &=& \half ( 1 \pm \gamma_0 )
\label{Lambda_0}
\ee
is the usual projector on positive and negative energy components. The
first power of the Foldy--Wout\-huysen matrices, $S^{\pm 1} (\bfq )$, can
be expressed from eq.(\ref{S2}) as
\be
S^{\pm 1}({\bf q}) &=& \cos\nu (\magq ) \,\pm\, \bfqh\!\cdot\!\vec{\gamma}
\sin \nu (\magq ) ,
\hspace{2em}
\nu (\magq ) \; =\; \half (\half\pi - \phi (\magq ) ) .
\label{foldy}
\ee
Inserting the quark propagator in the form of eq.(\ref{G_Sigma}) into the
Schwinger--Dyson equation, eq.(\ref{sde_eta}), and taking traces one
obtains a system of equations for $E(\magp )$ and $\phi (\magp )$,
\be
\begin{array} {rclcl}
E(\magp ) \sin\phi (\magp ) &=&  m^0 &-&
{\displaystyle \half \mint{q} \Vpq \sin\phi (\magq )}, \\
E(\magp ) \cos\phi (\magp ) &=& \magp &-&
{\displaystyle \half \mint{q} \Vpq \bfph\!\cdot\!\bfqh \cos\phi (\magq )} .
\end{array}
\label{sde_system}
\ee
These equations define the single-particle spectrum of two quarks forming a
bound state. From eqs.(\ref{sde_system}) one obtains an integral equation
for the chiral angle,
\be
\lefteqn{m^0 \cos\phi (\magp )  - \magp \sin\phi (\magp )
\; = \; } \nonumber \\
&& \half\mint{q}\Vpq\,\left[
\cos\phi (\magp ) \sin\phi (\magq ) -  \bfph\!\cdot\!\bfqh\,\sin\phi (\magp )
\cos\phi (\magq ) \right] .
\label{sde_phi}
\ee
Here, $\phi (\magp )$ satisfies the boundary conditions $\phi (0) =
\half\pi$ and $\phi (\magp ) \rightarrow 0$ for $\magp \rightarrow \infty$.
{}From the chiral angle, the quark energy can be obtained as
\be
\lefteqn{E (\magp ) \; =\; m^0 \sin\phi (\magp ) + \magp \cos\phi (\magp )}
\nonumber \\
&& - \half\mint{q}\Vpq\,\left[
\sin\phi (\magp ) \sin\phi (\magq ) +  \bfph\!\cdot\!\bfqh\,\cos\phi (\magp )
\cos\phi (\magq) \right] .
\label{sde_energy}
\ee
We remark that absolute values of the quark energy may be shifted by a
constant amount, $E_0$, without affecting the bound state spectrum, by
adding to the potential a contact term, $V_0 (\bfp - \bfq ) = - E_0 (2\pi
)^3 \delta^{(3)} (\bfp - \bfq )$, \cf\ the discussion in
\cite{adler_84,leyaouanc_84}.
\par
If the potential contains a Coulomb interaction, $V_{\rm C}(\bfp - \bfq ) =
- \frac{4}{3}\alpha_s/(\bfp - \bfq )^2$, the integrals in
eqs.(\ref{sde_system}) are ultraviolet divergent and require
renormalization. Consequently, we perform in eqs.(\ref{sde_system}) a wave
function and mass renormalization and replace the bare quark kinetic energy
and mass, $\magp$ and $m^0$, by
\be
\magp &\rightarrow& \magp \; +\; \half \mint{q} V_{\rm C}(\bfp - \bfq )\,
\bfph\!\cdot\!\bfqh ,
\label{renorm_p} \\
m^0 &\rightarrow& m^0 \; +\; \half \mint{q} V_{\rm C}(\bfp - \bfq )
\frac{m^0}{\sqrt{\magq^2 + m^{0\, 2}}} .
\label{renorm_m}
\ee
The equations resulting from eqs.(\ref{sde_system}, \ref{sde_phi},
\ref{sde_energy}) upon this substitution are finite. The boundary
conditions on the chiral angle remain unchanged. Furthermore, with the
definitions eqs.(\ref{renorm_p}, \ref{renorm_m}) the single--particle
energy for $\magp \rightarrow \infty$ behaves like that of a free quark. In
the case of an asymptotically free potential, $\alpha_s = \alpha_s (\magp
)$ it has been shown that the Schwinger--Dyson equation can be renormalized
consistently by requiring that the quark propagator reduce to the free
propagator at some large 3--momentum $\bfp_{\rm ren}$, with $|\bfp_{\rm
ren}| \rightarrow \infty$ \cite{adler_84,alkofer_88}. Here, we are working
with a potential, the UV--divergent part of which is of pure Coulomb form,
$\alpha_s = {\rm const.}$, so that the definition eq.(\ref{renorm_p}) is
sufficient\footnote{For a pure Coulomb potential without a mass scale the
definition eq.(\ref{renorm_p}) falls within the scope of multiplicative
renormalization. In other words, for a pure Coulomb potential the
subtraction procedure of Finger and Mandula \cite{finger_82} coincides with
the renormalization of Adler and Davis \cite{adler_84}.}. The linear
potential in eq.(\ref{potential}) is UV--finite and does not contribute to
the renormalization. The redefinition of $m^0$, eq.(\ref{renorm_m}), is not
a renormalization in the strict sense, as it corresponds to using a
momentum--dependent renormalization constant. In the framework of this
potential model we take the point of view that eq.(\ref{renorm_m}) is
simply a prescription to softly break chiral symmetry for the light flavors
with parameter $m^0$. Hence the value of $m^0$ is not inferred from the QCD
current mass but may be determined phenomenologically {\em e.g.} by fitting
the pion mass. This is usual practice in the framework of the
Nambu--Jona-Lasinio model \cite{volkov_83,reinhardt_86}. Definition
eq.(\ref{renorm_m}) leads to a smooth behavior of the pion mass and decay
constant in the chiral limit, as can be seen from figs.\ 4 and 5. For heavy
quarks, we will use the approximation of a constant constituent quark mass,
\cf\ below.
\par
The ratio of the scalar to vector part of the quark self--energy,
\be
m(\magp ) &=& \frac{A(\magp )\,\magp}{1 + B(\magp )} \; = \; \magp
\tan\phi (\magp ) ,
\label{dynamical_mass}
\ee
can be interpreted as a momentum--dependent ``constituent'' quark mass,
which reduces to the current quark mass, $m^0$, in the limit
$\magp\rightarrow\infty$ \cite{alkofer_88}.
\par
Eq.(\ref{sde_phi}) constitutes a non-linear integral equation for the
chiral angle, $\phi (\magp )$. By performing the angular integral one
obtains a 1--dimensional equation, which we solve numerically using the
relaxation methods of Adler \etal\ \cite{adler_84,piran_84}.
\par
Numerical solutions of the Schwinger--Dyson equation for the chiral angle,
$\phi (\magp )$, and the quark energy, $E(\magp )$, for a light quark
flavor are shown in figs.\ 1 and 2. Here, the potential is the one used
below to describe $D$-- and $B$--mesons ($\sigma = 0.41\,{\rm GeV},
\alpha_s = 0.39$). Note that the attractive interaction causes the quark
energy to become negative at small momenta. This fact is intimately related
to the spontaneous breaking of chiral symmetry and the emergence of
Goldstone bosons \cite{leyaouanc_84}. The dynamical quark mass, $m (\magp
)$, is shown in fig.3. We also show the corresponding quantities for the
oscillator model of \cite{leyaouanc_84} ($V_0 = 0.247\,{\rm GeV}$), for
which eq.(\ref{sde_phi}) reduces to a differential equation for $\phi
(\magp )$.
\par
For heavy quark flavors, the bare quark mass, $m^0$, is much larger than
the dynamically generated mass. One may then neglect the dynamical
contribution to the mass and approximate $\Sigma (\bfq )$ by a constant
constituent quark mass, $\Sigma ({\bf p}) \equiv m_Q$. In this case we
have in eq.(\ref{A_B}) $A(\magp ) \magp = m_Q ,\; B(\magp ) = 0$, so that
\be
\phi (\magp ) &=& \arctan\frac{m_Q}{\magp} , \hspace{1.8em}
E (\magp ) \; =\; \sqrt{\magp^{2} + m_Q^{2}} , \hspace{1.8em}
\nu (\magp ) \; = \; \half \arctan\frac{\magp}{m_Q} . \hspace{1.5em}
\label{constant_mass}
\ee
The chiral angle and the quark energy for a heavy quark are also shown in
figs.\ 1 and 2. A constant quark mass is also frequently used to describe
the light--quark sector, see {\em e.g.}\ \cite{resag_93}. We will comment
on the consequences of this approximation below.
\par
In the heavy--quark limit, if $m_Q$ becomes much larger than the momentum
of the relative motion of the heavy--light bound state, one may approximate
the Foldy--Wouthuysen angle of the heavy flavor by its value at $|\bfp | =
0$, $\phi (\magp ) \equiv \half\pi , \,\nu (\magp ) \equiv 0$.
Furthermore, the heavy--quark energy becomes $E(\bfp ) \sim m_Q + \half
\magp^2 /m_Q \sim m_Q + O(1/m_Q )$; to leading order one can neglect the
heavy--quark kinetic energy. In other words, in this approach the
heavy--quark limit amounts to neglecting the variations of the heavy--quark
energy and chiral angle over the range of momenta contributing to the
bound--state wave function.
\subsection{Equations for bound state wave functions}
Given the quark single--particle spectrum in the vacuum of the effective
theory, we now turn to the description of meson bound states. For the
covariantly written potential kernel, eq.(\ref{K_eta}), the Bethe--Salpeter
amplitude, $\Gamma (p|\cP )$, depends only on the transverse part of the
internal momentum, $p^\perp$. With this interaction the Bethe--Salpeter
equation, eq.(\ref{bse}), for a bound state of flavor $q_i \bar{q}_j$ in an
arbitrary frame reads
\be
\Gamma(p^\perp \vert \cP ) &=& i \int \frac{d^{4}q}{(2\pi)^{4}}
V(p^\perp - q^\perp )\, \rlap/{\eta} G_i ( q + \half\cP ) \Gamma (q^\perp
\vert \cP ) G_j ( q - \half\cP ) \rlap/{\eta} .
\label{bse_eta}
\ee
Here, $G_{n} \equiv G_{\Sigma_{n}}$ is the quark propagator of flavor $n$
defined in eq.(\ref{G_Sigma}), and $\cal{P}$ denotes the total momentum of
the bound state. For an instantaneous interaction it is convenient to
describe the bound state in terms of a Salpeter wave function,
\be
\Psi (q^\perp \vert \cP) &=&
i \int \frac{dq^{\vert\vert}}{2\pi} G_{i}(q + \half\cP )
\Gamma ({\bf q} \vert \cP ) G_{j}(q - \half\cP ) .
\label{Psi_dressed_general}
\ee
where $q^{\vert\vert} = q\!\cdot\!\eta$. In the following we shall restrict
ourselves to the rest frame, where $\cP = (M, \vec{0})$, with $M$ the bound
state mass, and $q^{\vert\vert} = q_0$. We drop the argument $\cP$ in
$\Gamma (\bfp |\cP ), \Psi (\bfp |\cP )$ in the following. Performing in
eq.(\ref{bse_eta}) the integral over $q_0$ one obtains for $\Psi (\bfp )$
the equation
\be
\Psi(\bfp ) &=& \frac{\Pi_{+-}(\bfp )}{E_p - M}
          \, +\, \frac{\Pi_{-+}(\bfp )}{E_p + M} .
\label{salpeter_eq_dressed}
\ee
Here, $E_p = E_i (\magp ) + E_j (\magp )$ is the sum of the
single--particle energies for the two quark flavors, and
\be
\Pi_{\pm\mp}(\bfp ) &=& \Lambda_{i\pm}(\bfp ) \gamma_{0}
\Gamma (\bfp ) \gamma_{0} \olin{\Lambda}_{j\mp}(\bfp ) , \\
\Gamma (\bfp ) &=& \mint{q} \Vpq \gamma_{0} \Psi(\bfq ) \gamma_{0} ,
\ee
with the projectors $\Lambda_{n\pm}(\bfp ), \olin{\Lambda}_{n\mp}(\bfp )$
for flavor $n$ defined in eq.(\ref{projectors}). Rather than working with
eq.(\ref{salpeter_eq_dressed}) directly it is more suitable to introduce an
``undressed'' wave function,
\be
\Psiz (\bfp ) &=& S_i (\bfp )\Psi(\bfp ) S_j (\bfp ) .
\label{Psi_undressed}
\ee
Here, $S_n (\bfp )$ are the Foldy--Wouthuysen matrices, eq.(\ref{foldy}),
corresponding to the two quark flavors. The new wave function satisfies an
equation analogous to eq.(\ref{salpeter_eq_dressed}),
\be
\Psiz (\bfp ) &=& \frac{\stackrel{0}{\Pi}_{+-}(\bfp )}
{E_p - M} \, + \, \frac{\stackrel{0}{\Pi}_{-+}(\bfp )}{E_p + M} ,
\label{salpeter_eq_undressed}
\ee
where $\stackrel{0}{\Pi}_{\pm\mp}(\bfp )$ now involves the free projectors,
eq.(\ref{Lambda_0}),
\be
\stackrel{0}{\Pi}_{\pm\mp}(\bfp ) &=& - \Lambdaz_{\pm}
S_i^{-1}(\bfp ) \Gamma(\bfp ) S_j^{-1}(\bfp ) \Lambdaz_{\mp} .
\label{Pi_0}
\ee
As a consequence, the new wave function satisfies the constraints
\be
\Lambdaz_\pm\Psiz (\bfp )\Lambdaz_\pm &=& 0 .
\label{constraint}
\ee
By applying projection operators, eq.(\ref{salpeter_eq_undressed}) can be
written in the form of a Schr\"odinger equation,
\be
[E_p \mp M]\Lambdaz_{\pm}\Psiz(\bfp )
\Lambdaz_{\mp} &=& \stackrel{0}{\Pi}_{\pm\mp}(\bfp ) \,\,.
\label{schroedinger_eq}
\ee
It has the structure of a generalized eigenvalue equation for the bound
state wave function, $\Psiz(\bfp )$, with $M$ as eigenvalue\footnote{The
mathematical structure of the instantaneous Bethe--Salpeter equation has
recently been analyzed in \cite{resag_93,lagae_92}.}. We now decompose
$\Psiz(\bfp )$ as
\be
\Psiz (\bfp ) &=& \Psiz^{(1)}(\bfp )
+ \, \gamma_{0} \cdot \Psiz^{(2)}(\bfp ) .
\label{Psi_decomposition}
\ee
Because of the constraint, eq.(\ref{constraint}), we can expand the
functions $\Psiz^{(k)}(\bfp )$ completely in the set of Dirac matrices
$\{\gamma_{5},\, i\vec{\gamma} \}$. We write
\be
\Psiz^{(k)}(\bfp )&=& L^{(k)}(\bfp)\,\gamma_{5}\, +\, i\bfN^{(k)}(\bfp )
\cdot\vec{\gamma} \hspace{2em} (k = 1, 2).
\label{Psi_L_N}
\ee
By taking traces of eq.(\ref{schroedinger_eq}) one obtains after a
straightforward but lengthy calculation a system of equations for the
component functions, $L^{(k)}(\bfp ), \bfN^{(k)} (\bfp ), \, k = 1, 2$. In
calculating the traces it is convenient to introduce a dreibein in momentum
space, $\{ \bfph ,\, \hat{e}^1_\bfp ,\, \hat{e}^2_\bfp \}$, with
$\bfph\!\cdot\!\hat{e}^a_\bfp = 0,\;
\hat{e}^{a}_\bfp\!\cdot\!\hat{e}^{b}_\bfp = \delta^{ab}\; (a, b = 1, 2)$
and expand the wave function in the set $\{\gamma_{5},\,
i\hat{e}^{a}_\bfp\!\cdot\!\vec{\gamma},\, i\bfph\!\cdot\!\vec{\gamma} \}$.
We find
\be
\lefteqn{M \twocomp{L}{2}{1}(\bfp ) \; - \; E_p \twocomp{L}{1}{2}(\bfp )
\; = \;
\mint{q} \Vpq} \hspace{4em}
\label{L_eq} \\
&& \times\Bigl\{ [c^\mp_p c^\mp_q \, +\, s^\mp_p s^\mp_q\,\bfph\!\cdot\!\bfqh ]
\,\twocomp{L}{1}{2}(\bfq )
\; + \; s^\mp_p s^\pm_q\, \bfph\!\cdot\!
(\bfqh\times \twocomp{\bfN}{2}{1}(\bfq ) ) \Bigr\}
\nonumber \\
\lefteqn{M \twocomp{\bfN}{2}{1} (\bfp ) \; - \; E_p
\twocomp{\bfN}{1}{2}(\bfp ) \; = \; \mint{q} \Vpq} \hspace{4em}
\label{N_eq} \\
&& \times\Bigl\{
 [ c^\mp_p c^\mp_q\, P^T_\bfp P^T_\bfq + c^\mp_p c^\pm_q\, P^T_\bfp
P^L_\bfq
 + c^\pm_p c^\mp_q\, P^L_\bfp P^T_\bfq + c^\pm_p c^\pm_q\, P^L_\bfp
P^L_\bfq
\nonumber \\
&& \hspace{1em} - s^\mp_p s^\mp_q\, \bfph\times (\bfqh\times \, .\, )
 + s^\pm_p s^\pm_q\, \bfph\, (\bfqh\!\cdot \, . \, ) ]\,
\twocomp{\bfN}{1}{2} (\bfq )
\nonumber \\
&& \hspace{1em} - s^\pm_p s^\mp_q\, \bfph\times\bfqh\,\twocomp{L}{2}{1}(\bfq )
\Bigr\}
\nonumber
\ee
Here, $P^T_\bfp , P^L_\bfp$ are the transverse and longitudinal
3--dimensional projectors,
\be
P^L_\bfp &=& \bfph\otimes\bfph , \hspace{2em}
P^T_\bfp \; =\; 1 - P^L_\bfp\; =\; -\bfph\times (\bfph\times .\, ) ,
\ee
with a similar definition for $\bfqh$. The factors $c^\pm_p , s^\pm_p$ are
the result of the Foldy--Wout\-huy\-sen transformation,
eq.(\ref{Psi_undressed}). We have introduced the short--hand notation
\be
c^{\pm}_p &=& \cos\nu_{ip}\cos\nu_{jp} \mp \sin\nu_{ip}\sin\nu_{jp}
\; = \; \cos (\nu_{ip} \pm \nu_{jp}) ,
\label{c_s} \\
s^{\pm}_p &=& \sin\nu_{ip}\cos\nu_{jp} \pm \cos\nu_{ip}\sin\nu_{jp}
\; = \; \sin (\nu_{ip} \pm \nu_{jp}) ,
\nonumber
\ee
with a similar definition for $c^\pm_q , s^\pm_q$. The angle $\nu_{np}
\equiv \nu_n (\magp )$ for flavor $n = i, j$ has been defined in
eq.(\ref{foldy}). The functions $E_p = E_i (\magp ) + E_j (\magp )$ and
$\nu_{ip}, \nu_{jp}$ contain the entire information on the single--particle
spectrum. They have to be provided either as solutions of the
Schwinger--Dyson equation for the quark self energy, eq.(\ref{sde_system}),
or by the approximation of a constant quark mass, eq.(\ref{constant_mass}).
\par
The equations eqs.(\ref{L_eq}, \ref{N_eq}) for the bound state wave
functions have a very compact form. Note in particular that the
pseudoscalar--axial and scalar--vector meson mixing is taken into account
here. Corresponding equations have been given by Le Yaouanc \etal\ for the
case of equal quark masses and a harmonic oscillator potential
\cite{leyaouanc_85}. Through the use of the Foldy--Wouthuysen
transformation, eq.(\ref{Psi_undressed}), we have preserved the simple
structure of the wave function, eq.(\ref{constraint}), even in the general
case of unequal quark flavors. Note that in the case of identical quark
flavors (isospin limit), one has $s^-_p = 0$ in eq.(\ref{c_s}), and the
equations for the $L$-- and ${\bf N}$--component decouple\footnote{The
equations for wave functions written down in \cite{kalin_90} apply only to
this case of equal quark flavors, since the $L$--$\bfN$ mixing term has
been omitted there.}. We remark that equations in a parametrization
somewhat different from to that of eqs.(\ref{L_eq}, \ref{N_eq}) have been
derived by Laga\"e using a variational approach \cite{lagae_92}. The use of
the Foldy--Wouthuysen transformation in the form eq.(\ref{Psi_undressed})
provides a simple and more transparent alternative.
\par
The bound state wave functions are normalized by condition
eq.(\ref{L_N_normalization}), which fixes the relativistic dispersion law
for the bound state. From eqs.(\ref{Psi_undressed}, \ref{Pi_0},
\ref{schroedinger_eq}) the Bethe--Salpeter amplitude can be reconstructed
from the ``undressed'' wave function in the rest frame. The wave functions
contain the full information about the bound state on its mass shell. In
appendix B, the pseudoscalar meson decay matrix element is evaluated in
terms of the wave function.
\par
For calculation of the meson spectrum it is necessary to decompose
eqs.(\ref{L_eq}, \ref{N_eq}) into equations for bound states of given
angular momentum and parity in the rest frame. This is done in appendix
A.1. The resulting radial equations are solved numerically using the
Multhopp method \cite{boukraa_89}. In the presence of the
momentum--dependent Foldy--Wouthuysen factors in the integral equation this
momentum--space method is much more convenient than the commonly used
matrix methods, in which the matrix elements of the potential are evaluated
in position space \cite{resag_93,lagae_92}. In appendix A.2 we show how the
Multhopp method can be adapted so that essentially no complications arise
from the Foldy--Wouthuysen factors.
\section{Light--light and heavy--light mesons}
The bound state equations derived from the bilocal effective action,
eq.(\ref{W_eff}), describe mesons for arbitrary quark masses. We now apply
this description to the study of light--light and heavy--light mesons.
Before embarking on the numerical solution of the Salpeter equations it is
worthwhile to investigate some limiting cases. In particular, we wish to
demonstrate how eqs.(\ref{L_eq}, \ref{N_eq}) describe both pseudoscalar
Goldstone bosons in the chiral limit, $m^0 \rightarrow 0$, and degenerate
heavy--light mesons in the heavy--quark limit. Let us first consider
eq.(\ref{L_eq}) in the isospin limit with two light quark flavors,
\ie , for the pion. In this case one has $\nu_{ip} = \nu_{jp} \equiv
\nu (\magp ) = \half (\half\pi - \phi (\magp ))$, so that
\be
\ba{lcl}
c^+_p &=& \sin\phi (\magp ) \\
c^-_p &=& 1 \\
\ea
\hspace{3em}
\ba{lcl}
s^+_p &=& \cos\phi (\magp ) \\
s^-_p &=& 0
\ea
\ee
with $\phi (\magp )$ the solution of eq.(\ref{sde_system}). If furthermore
$m^0 = 0$, comparison of the Salpeter equation, eq.(\ref{L_eq}), with the
Schwinger--Dyson equation, eq.(\ref{sde_system}), shows that
eq.(\ref{L_eq}) possesses a massless Goldstone mode solution of the form
\be
L^{(1)}(\bfp ) &\propto& \sin\phi (\magp ),\hspace{3em}
L^{(2)}(\bfp )\; =\; 0, \hspace{3em}
M\; =\; 0 .
\ee
For this property it is crucial that chiral symmetry is broken
spontaneously by the same interactions which bind the quarks in the meson,
\ie , that the quark self--energy is taken as the solution of the
Schwinger--Dyson equation and not approximated by a constant constituent
quark mass \cite{muenz_93}. Note also that the pseudoscalar wave function
is normalizable only for $m^0 > 0$.
\par
Of particular interest is the case of a small but finite current quark
mass, the chiral limit. Specifically, we want to see how the pseudoscalar
meson mass and decay constant behaves in the limit $m^0\rightarrow 0$. We
thus solve the Schwinger--Dyson equation, eq.(\ref{sde_system}), and the
pseudoscalar bound state equation, eqs.(\ref{L_eq}, \ref{rad_1}), for small
current quark masses, $m_u^0 = m_d^0 = m^0$. Results for the linear plus
Coulomb potential as well as the oscillator model are shown in figs.\ 4 and
5. As can be seen from fig.4, the pion mass vanishes in the chiral limit
like $m_\pi^2 \propto m^0$, in accordance with current algebra. The pion
decay constant, $f_\pi$, is shown in fig.5. As expected, it approaches a
finite limit if $m^0 \rightarrow 0$. We remark that for a pure oscillator
potential ($V_0 = 0.247\,{\rm GeV}$) we reproduce in the chiral limit the
value of Le Yaouanc \etal , $f_\pi = 20\,{\rm MeV}$, while for a pure
linear potential ($\sigma = 0.40\,{\rm GeV}$) we find $f_\pi = 0.11\,{\rm
MeV}$, which is in agreement with the value obtained by Adler and Davis
\cite{adler_84}. Pion properties in the chiral limit have also been studied
by Alkofer and Amundsen using the inhomogeneous Bethe--Salpeter equation
\cite{alkofer_88}. For an instantaneous interaction, the wave function
description of bound states, eq.(\ref{Psi_dressed_general}), is a
convenient alternative, especially for the study of excited states. Also
shown in fig.5 is the decay constant for the radially excited pion state,
which vanishes in the chiral limit, as expected on general grounds
\cite{leyaouanc_85}. The reason for this is that as the pion wave function
approaches $\sin\phi (\magp )$, the excited state wave functions become
orthogonal to $\sin\phi (\magp )$ and thus the integral for the pion decay
constant, eq.(\ref{fpi}), vanishes. The pion radial wave function, \cf\
eq.(\ref{partial_waves}), is shown in fig.6, for a current mass of $m^0 =
1\,{\rm MeV}$.
\par
It is well--known that a charmonium (linear plus Coulomb) potential with
usual parameters underestimates the strength of spontaneous chiral symmetry
breaking, which manifests itself in the too small value of $f_\pi$
\cite{adler_84}. Improved values can be obtained by including either an
intermediate--range attractive potential \cite{hirata_89} or transverse
gluon exchange \cite{alkofer_88}.
\par
Thus, we have verified that the bilocal effective meson action,
eq.(\ref{W_eff}), reproduces the successful phenomenology of the pion as a
Goldstone boson, if the breaking of chiral symmetry is described consistently
with the interactions which form the bound state.
\par
Let us now consider heavy--light mesons. Specifically, we shall investigate
a bound state of a heavy quark and a light antiquark ($Q_i\bar{q}_j$). For
the heavy quark we neglect spontaneous chiral symmetry breaking and
approximate its energy by a constant mass, \cf\ eq.(\ref{constant_mass}).
In order to take the limit $m_Q\rightarrow\infty$, we subtract from $E_p$
and $M$ in eqs.(\ref{L_eq}, \ref{N_eq}) the heavy quark mass,
\ie , we consider binding energies relative to the heavy quark
mass. For the light flavor, we take take into account the dynamical
breaking of chiral symmetry and take $E_j (\magp ),\, \nu_j (\magp )$ as
the solution of the Schwinger--Dyson equation, eq.(\ref{sde_system}).
\par
In particular, in the heavy--quark limit, if $m_Q$ becomes much larger than
the range of momenta contributing to the integrals over the bound--state
wave function we may simply take $\nu_i (\magp ) \equiv 0$. In this case we
have
\be
\ba{rcrcrcr}
c^\pm_p &=& \cos(\pm\nu_{jp}) &=& \cos(\nu_{jp} ) &\equiv& c_p , \\
s^\pm_p &=& \sin(\pm\nu_{jp}) &=& \pm \sin(\nu_{jp} ) &\equiv& \pm s_p ,
\label{c_s_heavy}
\ea
\ee
while $E_i (\magp ) = \sqrt{m_Q^2 + \magp^2} = m_Q + O(1/m_Q )$. Consequently,
in eqs.(\ref{L_eq}, \ref{N_eq}) the equations for the $(1)$-- and
$(2)$--components of the wave function coincide, and the solutions satisfy
\be
L^{(1)}(\bfp ) &=& L^{(2)}(\bfp ) \; \equiv \; L(\bfp ), \hspace{2em}
\bfN^{(1)}(\bfp ) \; = \; \bfN^{(2)}(\bfp ) \; \equiv \; \bfN (\bfp ) .
\label{L1_L2_heavy}
\ee
Eqs.(\ref{L_eq}, \ref{N_eq}) thus simplify to
\be
\lefteqn{(M - E_p ) L(\bfp ) \; =\; \mint{q} \Vpq} \hspace{4em}
\label{L_heavy} \\
&&\times\Bigl\{ [ c_p c_q + s_p s_q\,\bfph\!\cdot\!\bfqh ]
\, L(\bfq )\; -\; s_p s_q\,\bfph\!\cdot\! (\bfqh\times\bfN (\bfq )) \Bigr\}
\nonumber \\
\lefteqn{(M - E_p ) \bfN (\bfp ) = \mint{q} \Vpq} \hspace{4em}
\label{N_heavy} \\
&&\times\Bigl\{ [ c_p c_q + s_p s_q
( (\bfph\!\cdot\!\bfqh) -  (\bfph\times \bfqh ) \times\, .\, )) ]\,\bfN (\bfq )
\; +\; s_p s_q \,\bfph\times\bfqh\, L (\bfq ) \Bigr\}
\nonumber
\ee
The reduced equations eqs.(\ref{L_heavy}, \ref{N_heavy}) exibit the
heavy--quark spin symmetry, \ie , the pseudoscalar bound state
is degenerate with the vector, the scalar with the axial vector
\cite{isgur_89}. This symmetry is a consequence of the fact that
the dynamics becomes independent of the heavy quark spin in the limit $m_Q
\rightarrow \infty$. We can demonstrate this explicitly for S--wave bound
states ($J^P = 0^-, 1^-$). In this case, the equation for the $L$-- and
$\bfN$--component decouple, and one may verify that if eq.(\ref{L_heavy})
possesses a pseudoscalar solution of the form $L(\bfq ) = f(|\bfq |)$,
eq.(\ref{N_heavy}) admits a vector solution corresponding to the same
orbital wave function multiplied by a constant polarization vector, $\bfN
(\bfp ) = \vec{S} f(|\bfp |) , \; \vec{S} = {\rm const.}$. (Here, we
suppose that the 3--dimensional potential is a function of $|\bfp - \bfq |$
only.) A similar degeneracy holds for arbitrary angular momentum and both
parities, as is seen from the corresponding partial--wave equations,
eqs.(\ref{rad_1_heavy}, \ref{rad_2_heavy}) in appendix A.1. Thus, the bound
state equations eqs.(\ref{L_eq}, \ref{N_eq}) naturally realize the
heavy--quark spin symmetry in the limit $m_Q \rightarrow \infty$.
\par
Let us briefly discuss the non--relativistic limit of this description of
mesons. In this case we take in eq.(\ref{c_s_heavy}) also $\nu_j (\magp )
\equiv 0$ and obtain $c^\pm_p \equiv 1, s^\pm_p \equiv 0$. In addition, we
approximate both quark energies by their non-relativistic values, $E_n
(\magp ) = m_n + \magp^2/2 m_n ,\, n = i, j$. In this limit,
eqs.(\ref{L_heavy}, \ref{N_heavy}) reduce to the well-known
non-relativistic potential model for heavy quarkonia \cite{eichten_78}.
\par
We have seen that the bilocal effective theory with instantaneous
interaction provides a unified description of both light and heavy flavors.
It should thus be well suited to investigate the spectrum and decays of
heavy--light bound states. Our intention here is not to perform an
exhaustive calculation of the meson spectrum based on the Salpeter
equations, eqs.(\ref{L_eq}, \ref{N_eq}), but rather to obtain a
quantitative estimate of the effects of dynamical chiral symmetry breaking
on the heavy--light meson spectrum and decay constants. Specifically, we
wish to demonstrate the phenomenological importance of describing chiral
symmetry breaking self--consistently through the Schwinger--Dyson equation,
eq.(\ref{sde_system}), rather than by a constant mass for the light quark,
eq.(\ref{constant_mass}). To this end, we compare the masses and decay
constants of the $D$-- and $B$--mesons calculated using the
momentum--dependent quark self energy from eq.(\ref{sde_system}) with those
obtained using a constant light quark mass, eq.(\ref{constant_mass}).
\par
To describe the $D$-- and $B$--mesons, we take as potential the sum of a
Coulomb and a linear potential, eq.(\ref{potential}), as has been used in
the analysis of charmonium, $c\bar c$ \cite{eichten_78}. In our
investigations here we use the same potential for $D$-- and $B$--mesons and
ignore the running of $\alpha_s$. For the parameters we take the values
$\sigma = 0.41\,{\rm GeV}$ and $\alpha_s = 0.39$, which we determined by
fitting the masses of the $1S$-- and $2S$--state of $c\bar c$ ($J/\psi$) as
well as those of $b\bar b$ ($\Upsilon$), using eqs.(\ref{L_eq}, \ref{N_eq})
in the non-relativistic limit, $\nu_{ip} = \nu_{jp} \equiv 0, E_i (\magp )
= E_j (\magp ) = m_c + \half \magp^2/ m_c$. In this fit the quark masses
were obtained as $m_c = 1.39\,{\rm GeV}$ and $m_b = 4.79\,{\rm GeV}$.
\par
The result of a calculation of the masses and decay constants of
pseudoscalar $D$-- and $B$--mesons and their first radially excited states
with the above potential is shown in table \ref{tab_data}. There, results
are given obtained with both the dynamical self--energy for the light
flavor and with a constant light quark mass. For the sake of comparison we
have chosen the constant light quark mass equal to the dynamically
generated quark mass, eq.(\ref{dynamical_mass}), at $\magp = 0$, which for
this potential is $m(0) =\, 0.082\, {\rm GeV}$, \cf\ fig.3. We have also
included values obtained with a typical light quark ``constituent'' mass,
$m_q = 0.33\, {\rm GeV}$. As can be seen, the use of the constant quark
mass instead of the dynamical self--energy leads to considerable changes in
the meson masses and decay constants. The increase in the decay constant of
the ground--state $D$-- and $B$--meson can be explained by the sensitivity
of $F_{D, B}$ to the higher--momentum part of the wave function, which is
larger if a constant quark mass is used. This is seen also from fig.7,
which shows the radial wave functions in both cases. Note further that the
$D$-- and $B$--meson masses calculated with the dynamical quark mass agree
rather well with the experimental values, much better than the ones
obtained with either $m_q =\, {\rm const.}$. (If $m_q$ is taken to be
significantly smaller than the values in table 1, the bound state mass even
increases.)  However, one should not overemphasize the quantitative
agreement, since we did not choose parameters such as to fit the strength
of chiral symmetry breaking in the light sector. More observables ($f_\pi
,\, \langle\bar q q\rangle$) should be involved in order to optimize the
potential before drawing conclusions based on absolute values.
Nevertheless, the results of table \ref{tab_data} show clearly that a
dynamical description of chiral symmetry breaking is important, and that
noticeable effects due to the dynamical nature of the quark masses occur
already in the meson mass spectrum and decay constants. Our results confirm
the conclusions of Kaburagi \etal , who investigated the dependence of
heavy--light meson properties on the light quark mass in the framework of
the Dirac equation with a scalar confining potential \cite{kaburagi_81}. It
should be stressed, however, that the main point in having a description
which takes in to account the dynamical breaking of chiral symmetry is a
qualitative one. The clearest manifestations of the role of dynamical
chiral symmetry in heavy--light systems may be seen not in the masses and
decay constants but in more complicated observables, {\em e.g.} processes
involving emission of pions. Such processes can be described using the
effective bilocal meson action, eq.(\ref{W_eff}).
\par
Finally, we would like to comment on the ordering of the pseudoscalar meson
decay constants. We find $F_B$ to be slightly larger than $F_D$, for both
the dynamical and constant light quark mass. This is in agreement with the
calculation of Cea \etal\ \cite{cea_88}, who use a relativistic potential
model, but contrary to most non-relativistic results, \cf\
\cite{markIII_88} and references therein. In \cite{cea_88} it is argued
that relativisitc effects spoil the simple proportionality $F_H \propto
M_H^{-1/2}$. If we treat in our approach the heavy quark as
non-relativistic, \ie , if we solve eqs.(\ref{L_eq}, \ref{N_eq}) with
$\nu_{1p} \equiv 0$ and $E_{1p} = m_Q + \half \magp^2 /m_Q$, we find $F_D$
to be somewhat larger than $F_B$ for both the dynamical self--energy and
the constant quark mass for the light flavor. Thus, the pattern is reversed
if the relativistic kinematics is abandoned. Note that our spectra were
calculated using the same potential for $D$-- and $B$--mesons; larger
differences in the decay constants may occur with a running coupling
constant.
\section{Summary and outlook}
In this paper we have presented a bilocal effective model, which describes
in a unified way both light and heavy mesons. Chiral symmetry is broken
spontaneously by the interactions which bind the mesons. The important
feature of our model is that the potential kernel, eq.(\ref{K_eta}), moves
together with the bound state, which leads to a relativistically covariant
description of bound states. The bilocal meson action obtained from this
model, eq.(\ref{W_eff}), provides the script for deriving equations for
bound states and for the calculation of matrix elements. The correct
relativistic kinematics is essential in describing meson decays and more
complicated processes like semileptonic decays \cite{hussain_90,weiss_94}.
This covariant formulation, which constitutes a relativistic extension of
the usual Coulomb gauge, should also be useful in the light quark sector,
for example in describing the pion electromagnetic form factor
\cite{langfeld_89} or processes like $\rho\rightarrow\pi\pi$.
\par
The Salpeter equations for the meson wave functions can be simplified
considerably by a Foldy--Wouthusen transformation,
eq.(\ref{Psi_undressed}). Of practical importance is the fact that this
transformation does not lead to complications in the numerical solution of
the bound state equations if the momentum--space Multhopp method is
employed. The equations exhibit both chiral symmetry and heavy--quark spin
symmetry in dependence on the chiral angle specifying the quark
single--particle spectrum.
\par
With a phenomenological potential used in the description of charmonium,
good results for the masses of the pseudoscalar $D$-- and $B$--mesons are
obtained if spontaneous chiral symmetry breaking is taken into account.
Heavy--light meson masses and decay constants are seen to be sensitive to
the momentum dependence of the light quark self--energy. This demonstrates
the necessity of formualting chiral symmetry breaking self--consistently
and justifies the effort of solving the Schwinger--Dyson equation in the
light flavor sector.
\par
Our aim here has been to set up the framework for describing heavy--light
meson bound states in a bilocal effective theory. We have tried to
demonstrate the viability of this scheme and to get a quantitative picture
of the meson spectrum. Clearly, there is much room to improve and extend
the results presented here. For example, an intermediate--range potential
or transverse gluon exchange could be included in addition to the Coulomb
and confinement potential in order to increase the strength of chiral
symmetry breaking in the light sector \cite{hirata_89,alkofer_88}.
Furthermore, an interesting possibility would be to consider the effective
meson action with the covariant instantaneous interaction in the
heavy--quark limit and perform a long--wavelength expansion, as outlined in
\cite{nowak_93}. This would allow one to make quantitative predictions for
the lagrangian of heavy quark effective theory in dependence on the
interaction potential assumed at quark level. Moreover, the covariant
instantaneous formulation could be generalized to the particle--particle
sector of the quark theory, eq.(\ref{action_q}), to describe also baryons
as bound states.
\vspace{1cm} \\
{\large\bf Acknowledgement} \\ This work was begun during the
Heisenberg--Landau Workshop ``Effective QCD at finite temperature'' at the
Joint Institute of Nuclear Research in Dubna, May 19 -- 23, 1993. Yu.\ L.
K. thanks the University at T\"ubingen, C. W. the University at Rostock and
the Joint Institute in Dubna for their hospitality.
%
%
\renewcommand{\theequation}{\thesection.\arabic{equation}}
\appendix
\newpage
\section{The Salpeter equation}
\setcounter{equation}{0}
\subsection{Partial wave decomposition}
In this section we decompose the Schr\"odinger--type equations in the rest
frame, eqs.(\ref{L_eq}, \ref{N_eq}), into equations for bound states of
given total angular momentum and parity (\cf\ also \cite{lagae_92}). This
is achieved by expanding the functions $L(\bfp ),\,\bfN (\bfp)$ in
spherical harmonics of total angular momentum, $J$. For the vector
component it is convenient to use instead of the usual vector spherical
harmonics the combinations
\be
\ba{lclcl}
\bfY_{JM}^1  &=& \bfph Y_{JM}  &=&
\alpha \bfY_{JJ-1M} - \beta \bfY_{JJ+1M} , \\[1ex]
\bfY_{JM}^2 &=& (J(J + 1))^{-1/2}\,\nabla_{\bfph} Y_{JM}  &=&
\beta \bfY_{JJ-1M} + \alpha \bfY_{JJ+1M} , \\[1ex]
\bfY_{JM}^3 &=& - i (J(J + 1))^{-1/2}\,\bfph\times\nabla_{\bfph} Y_{JM}
&=& \bfY_{JJM} ,
\ea
\ee
with
\be
\alpha &=& \sqrt{\frac{J}{2J + 1}}, \hspace{2em}
\beta\; =\; \sqrt{\frac{J + 1}{2J + 1}}, \hspace{2em}
\alpha^2 + \beta^2 \; =\; 1 .
\label{mu_nu}
\ee
The new functions $\bfY^\lambda_{JM}$ have simple transformation properties
under the operations $\bfph\times\bfY^\lambda_{JM}$ and
$\bfph\cdot\!\bfY^\lambda_{JM}$. They are orthogonal and normalized as the
usual vector spherical harmonics. We thus write the component functions of
eq.(\ref{Psi_L_N}) as\footnote{In this section, $p, q$ denote $\magp ,
\magq$.}
\be
L^{(k)} (\bfp ) &=& \frac{\ell^{(k)}_{J}(p)}{p} Y_{JM}(\bfph ) ,
\hspace{1.5em}
\bfN^{(k)} (\bfp ) \; =\; \sum_{\lambda = 1}^3
\frac{n^{(k)}_{\lambda J}(p)}{p} \bfY^\lambda_{JM} (\bfph )
\hspace{1.5em} (k = 1, 2). \hspace{1em}
\label{partial_waves}
\ee
For the angular matrix element of the potential kernel we use the
definition
\be
\frac{pq}{(2\pi )^3}
\int d\Omega_p \int d\Omega_q Y^*_{L^{'} M^{'}}(\bfph ) \Vpq Y_{LM} (\bfqh )
&=& v_L (p , q) \delta_{L L^{'}}\delta_{M M^{'}} . \label{v_l}
\ee
Note that the matrix element is independent of $M$, and that $v_L (p, q)$
is a symmetric functions of the radial variables $p, q$. The general
expressions for $v_L (p, q)$ corresponding to the power--like potentials
used in eq.(\ref{potential}) are given in table \ref{tab_potential}. From
the formula
\be
\frac{pq}{(2\pi )^3}
\int d\Omega_p \int d\Omega_q \bfY^*_{J^{'} L^{'} M^{'}}(\bfph ) \Vpq
\bfY_{JLM} (\bfqh ) &=& v_L (p , q)\, \delta_{J J^{'}}\delta_{L L^{'}}
\delta_{M M^{'}} \hspace{1em} \\
(L \; = \; J, J\pm 1) &&
\nonumber
\ee
we obtain the matrix elements
\be
\lefteqn{
\frac{pq}{(2\pi )^3}
\int d\Omega_p \int d\Omega_q \bfY^{\lambda *}_{JM}(\bfph ) \Vpq
\bfY^{\rho}_{JM} (\bfqh ) \; = } \hspace{3em} \nonumber \\
&& \left\{
\ba{lcll}
\alpha^2 v_{J - 1} + \beta^2 v_{J + 1} &\equiv& \olin{v}_J (p, q)
\hspace{1.5em} & (\lambda , \rho ) = (1, 1) \\
\beta^2 v_{J - 1} + \alpha^2 v_{J + 1} &\equiv& \olin{\olin{v}}_J (p, q) &
(\lambda , \rho ) = (2, 2) \\
v_J & & & (\lambda , \rho ) = (3, 3) \\
\alpha \beta (v_{J - 1} - v_{J + 1}) &\equiv& \wtil{v}_J (p, q) &
(\lambda , \rho ) = (1, 2), (2, 1) \\
\ea \right.
\label{vector_matel}
\ee
All other combinations vanish.
\par
With the above definitions it is straightforward to reduce eqs.(\ref{L_eq},
\ref{N_eq}) to the following two systems of radial equations,
\be
M \twocomp{\ell_J}{2}{1} (p) - E_p \twocomp{\ell_J}{1}{2} (p)
&=& \int_0^\infty dq\, \Bigl\{ [ c^\mp_p c^\mp_q v_J +
s^\mp_p s^\mp_q \olin{v}_J ] \twocomp{\ell_J}{1}{2} (q)
+ i s^\mp_p s^\pm_q \wtil{v}_J \twocomp{n}{2}{1}_{3J} (q) \Bigr\}
\nonumber \\
M \twocomp{n}{2}{1}_{3J} (p) - E_p \twocomp{n}{1}{2}_{3J} (p)
&=& \int_0^\infty dq\, \Bigl\{ [ c^\mp_p c^\mp_q v_J +
s^\mp_p s^\mp_q \olin{\olin{v}}_J ] \twocomp{n}{1}{2}_{3J} (q)
- i s^\mp_p s^\pm_q \wtil{v}_J \twocomp{\ell_J}{2}{1} (q) \Bigr\}
\nonumber \\
\label{rad_1}
\ee
and
\be
M \twocomp{n}{2}{1}_{1J} (p) - E_p \twocomp{n}{1}{2}_{1J} (p)
&=& \int_0^\infty dq\, \Bigl\{ [ c^\pm_p c^\pm_q \olin{v}_J +
s^\pm_p s^\pm_q v_J ] \twocomp{n}{1}{2}_{1J} (q)
+ c^\pm_p c^\mp_q \wtil{v}_J \twocomp{n}{1}{2}_{2J} (q) \Bigr\}
\nonumber \\
M \twocomp{n}{2}{1}_{2J} (p) - E_p \twocomp{n}{1}{2}_{2J} (p)
&=& \int_0^\infty dq\, \Bigl\{ [ c^\mp_p c^\mp_q \olin{\olin{v}}_J +
s^\mp_p s^\mp_q v_J ] \twocomp{n}{1}{2}_{2J} (q)
+ c^\mp_p c^\pm_q \wtil{v}_J \twocomp{n}{1}{2}_{1J} (q) \Bigr\}
\nonumber \\
\label{rad_2}
\ee
Here, $v_J \equiv v_J (p, q)$ {\em etc.}. The radial equations describe
bound states of parity $(-)^{J + 1}$ and $(-)^J$, respectively.
\par
In the heavy--quark limit, with the Foldy--Wouthuysen factors
given by eq.(\ref{c_s_heavy}), $c^\pm_p \equiv c_p, s^\pm_p \equiv \pm s_p$,
the systems of radial equations eqs.(\ref{rad_1}, \ref{rad_2}) simplify to
\be
(M - E_p ) \bmat{c} \ell_J \\ n_{3J} \emat (p)
&=& \int_0^\infty dq\, [ c_p c_q \left(\!\ba{lr} v_J & \\ & v_J \ea\!\right)
+ s_p s_q \bmat{rr} \olin{v}_J & -i\wtil{v}_J \\ i\wtil{v}_J &
\olin{\olin{v}}_J \emat ]
\bmat{c} \ell_J \\ n_{3J} \emat (q)
\nonumber \\
\label{rad_1_heavy} \\
(M - E_p ) \bmat{c} n_{1J} \\ n_{2J} \emat (p)
&=& \int_0^\infty dq\, [ c_p c_q \bmat{lr} \olin{v}_J & \wtil{v}_J \\
\wtil{v}_J & \olin{\olin{v}}_J \emat
+ s_p s_q \bmat{lr} v_J & \\ & v_J \emat ]
\bmat{c} n_{1J} \\ n_{2J} \emat (q)
\nonumber \\ \label{rad_2_heavy}
\ee
Here, we have dropped the indices on the wave function components in
accordance with eq.(\ref{L1_L2_heavy}). These reduced equations can be
diagonalized by introducing new radial wave functions,
\be
\ba{rcrcl}
a_{1J} &=& \alpha \ell_J &-& i\beta n_{3J} , \\
a_{2J} &=& -\beta \ell_J &-& i\alpha n_{3J} , \\
\ea
\hspace{3em}
\ba{rcrcl}
b_{1J} &=& \alpha n_{1J} &+& \beta n_{2J} , \\
b_{2J} &=& -\beta n_{1J} &+& \alpha n_{2J} . \\
\ea
\ee
For the transverse vector components this is equivalent to using in
eq.(\ref{partial_waves}) instead of the $\bfY^\lambda_{JM}$ the ordinary
vector spherical harmonics, $\bfY_{JLM}$, as basis functions. Upon this the
system eqs.(\ref{rad_1_heavy}, \ref{rad_2_heavy}) becomes
\be
(M - E_p )
\bmat{c} a_{1J} \\ a_{2J} \emat (p)
&=& \int_0^\infty dq\, [ c_p c_q \bmat{lr} v_J & \\ & v_J \emat
+ s_p s_q \bmat{lr} v_{J - 1} & \\ & v_{J + 1} \emat ]
\bmat{c} a_{1J} \\ a_{2J} \emat (q)
\nonumber \\ \label{rad_tf_1} \\
(M - E_p )
\bmat{c} b_{1J} \\ b_{2J} \emat (p)
&=& \int_0^\infty dq \, [ c_p c_q
\bmat{lr} v_{J - 1} & \\ & v_{J + 1} \emat
+ s_p s_q \bmat{lr} v_J & \\ & v_J \emat ]
\bmat{c} b_{1J} \\ b_{2J} \emat (q)
\nonumber \\ \label{rad_tf_2}
\ee
The new system eqs.(\ref{rad_tf_1}, \ref{rad_tf_2}) exhibits the
heavy--quark spin symmetry. For a given $J$ the equations for $a_{1, J +
1}$ and $a_{2, J}$ coincide with those for $b_{2, J}$ and $b_{1, J + 1}$,
respectively. This means that the bound states come in degenerate pairs
with total angular momentum $J$ and $J+1$, for both positive and negative
parity. For example, in this limit the $0^-$ (pseudoscalar) meson is
degenerate with the $1^-$ (vector), and the $0^+$ (scalar) meson with the
$1^+$ (axial vector).
\subsection{Numerical solution: the Multhopp method}
The radial equations, eqs.(\ref{rad_1}, \ref{rad_2}), are in general
singular integral equations with power--like or logarithmic singularities
at $q = p$. A simple and powerful numerical technique to solve such
equations directly in momentum space is the Multhopp method, which has been
used frequently in the context of non-relativistic and relativistic
constituent quark models \cite{boukraa_89}. Of crucial importance for this
method is the fact that for potentials with known singularities the
integral of the basis functions with the singular part of the potential
kernel can be split off and performed analytically, so that only finite
integrals need to be evaluated numerically. In the radial equations
eqs.(\ref{rad_1}, \ref{rad_2}) a new feature compared to the quark model is
the presence of the momentum--dependent Foldy--Wouthuysen factors, $c^\pm_q
, s^\pm_q$, modifying the integration kernel. Here, we briefly show how the
difficulties presented by these additional form factors can be circumvented
and the singular part of the integrals be separated as usual. Thus, the
Multhopp method can be used just as efficiently in the case of a
non-trivial single--particle spectrum as in standard quark model
calculations.
\par
For simplicity, we consider eq.(\ref{rad_1}) for a $J^P = 0^-$ bound state;
the generalization to the coupled equations for $J > 0$ is straightforward.
In this case $\wtil{v}_0 = 0, \, \olin{v}_0 = v_1$ and eq.(\ref{rad_1}) for
$\ell^{\,(1, 2)} (p) \equiv \ell^{\,(1, 2)}_0 (p)$ simplifies to
\be
M \twocomp{\ell}{2}{1} (p) - E_p \twocomp{\ell}{1}{2} (p)
&=& \int_0^\infty\! dq\, [ c^\mp_p c^\mp_q v_0 (p, q) +
s^\mp_p s^\mp_q v_1 (p, q) ] \twocomp{\ell}{1}{2} (q)
\label{multhopp_example}
\ee
The Multhopp method consists in converting eq.(\ref{multhopp_example}) into
a matrix equation in momentum space. The range of momenta $(0, \infty )$ is
mapped onto the interval $(0, \pi )$ by the coordinate transformation
\be
p &=& \lambda \tan\half\theta , \hspace{3em} q\; = \; \lambda
\tan\half\chi .
\label{p_theta}
\ee
where $\lambda$ is a scale parameter with dimensions of momentum. One then
chooses as basis functions the finite set $(2/\pi )^{1/2} \sin i\theta ,\;
i = 1,\ldots N$. The expansion of $\ell^{\, (1, 2)} (\theta )$ is
equivalent to interpolating $\ell^{\, (1, 2)} (\theta )$ at a set of angles
$\theta_k = k\pi/(N + 1), \; k = 1, \ldots N$, the so-called Multhopp
angles. By making use of the orthogonality relation for the basis
functions one obtains from eq.(\ref{multhopp_example}) a $2N\times
2N$--eigenvalue equation for $\ell^{\, (1, 2)} (\theta_k )$,
\be
\sum_{k = 1}^N \twocomp{B}{1}{2}_{jk} \twocomp{\ell}{1}{2} (\theta_k ) &=&
M \twocomp{\ell}{2}{1} (\theta_j ),
\label{multhopp}
\ee
with
\be
\twocomp{B}{1}{2}_{jk} &=& E(\theta_j ) \delta_{jk}
+ \frac{2}{N + 1} \sum_{i = 1}^N \sin i\theta_k
\twocomp{B}{1}{2}(i, \theta_j ), \\
\twocomp{B}{1}{2}(i, \theta_j ) &=& \int_0^\pi d\chi
\frac{\lambda}{2\cos^2 \half\chi} [ c^\mp_\theta c^\mp_\chi
v_0 (\theta, \chi ) + s^\mp_\theta s^\mp_\chi
v_1 (\theta , \chi ) ] \sin i\chi
\label{multhopp_integral}
\ee
Here, $\ell^{\, (1, 2)}(\theta ), E(\theta ), c^\mp_\theta, s^\mp_\theta,
v_0 (\theta , \chi ), v_1 (\theta , \chi )$ are related to the
corresponding functions in eq.(\ref{multhopp_example}) by the
transformation, eq.(\ref{p_theta}). The solution of eq.(\ref{multhopp})
directly yields the momentum--space wave function at a discrete set of
points \cite{boukraa_89}. The calculations quoted in section 4 were
performed with $N = 20\ldots 30$.
\par
The main input to eq.(\ref{multhopp}) are the Multhopp integrals of the
potential kernel, eq.(\ref{multhopp_integral}). In particular, the
singularity of the integration kernel at $q = p$ is absorbed in the
integral over the basis functions, eq.(\ref{multhopp_integral}). We now
show that by a simple rearrangement of terms the part of
eq.(\ref{multhopp_integral}) containing the singularity can be separated
even in the presence of the Foldy--Wouthuysen factors. The Coulomb
potential produces only an integrable logarithmic singularity in the radial
equation, \cf\ table \ref{tab_potential}, and poses no further problems.
For the linear potential the singularity is of the type
\be
v_L (p, q)_{\,{\rm lin}} &=& \frac{1}{(p - q)^2} \, + \,
\hbox{\rm log.\ div.\ terms} .
\label{singular_lin}
\ee
Note that the leading singularity is independent of the angular momentum,
$L$. Let us consider the contribution of the leading singularity,
eq.(\ref{singular_lin}), to the radial integral of
eq.(\ref{multhopp_example}). As the leading singularities of $v_0 (p, q)$
and $v_1 (p, q)$ are identical, the singular part of the integral in
eq.(\ref{multhopp_example}) is given by the principal value integral
\be
{\rm P}\int_0^\infty\frac{dq}{(q - p)^2}\,
[ c^\mp_p c^\mp_q + s^\mp_p s^\mp_q ] \twocomp{\ell}{1}{2} (q) .
\label{singular_contr}
\ee
Since the radial wave functions satisfy $\ell^{(1, 2)} (0) = \ell^{(1, 2)}
(\infty ) = 0$, eq.(\ref{singular_contr}) can be integrated by parts, which
gives
\be
{\rm P}\int_0^\infty \frac{dq}{(q - p)}\,\frac{d}{dq} \Bigl\{
[ c^\mp_p c^\mp_q + s^\mp_p s^\mp_q ] \twocomp{\ell}{1}{2} (q) \Bigr\} .
\ee
This can be rewritten as
\be
\lefteqn{ {\rm P}\int_0^\infty
\frac{dq}{(q - p)}\,\frac{d}{dq} \twocomp{\ell}{1}{2}(q)}
\label{rearranged}\\
&+& \int_0^\infty\frac{dq}{(q - p)}\,\Bigl\{
[ c^\mp_p c^\mp_q + s^\mp_p s^\mp_q - 1]
\frac{d}{dq}\twocomp{\ell}{1}{2}(q)
\, + \, [s^\mp_p c^\mp_q - c^\mp_p s^\mp_q ]
\twocomp{\ell}{1}{2}(q)\,\frac{d}{dq}(\nu_{1q} \mp \nu_{2q}) \Bigr\}
\nonumber
\ee
Of the integrals in eq.(\ref{rearranged}) only the first one is singular,
which does not involve the Foldy--Wouthuysen angle in the integrand, while
the second one is non-singular. Thus, the radial integrals involving the
linear potential can be rearranged such that the Foldy--Wouthuysen factors,
which are in general known only numerically, enter only in non-singular
integrals. In other words, the Foldy--Wouthuysen transformation does not
modify the leading singularity of the radial equation, as is to be
expected. After the substitution, eq.(\ref{p_theta}), the Multhopp
integral corresponding to the singular part of eq.(\ref{rearranged}) can be
performed analytically as usual \cite{boukraa_89}, because the
Foldy--Wouthuysen factors do not enter. The remaining integrals of
eq.(\ref{rearranged}), which involve the Foldy--Wouthuysen factors, lead to
Multhopp integrals with at most logarithmic singularities, which can be
calculated efficiently with standard Fourier transform routines.
\par
The above argument can easily be extended to the full equations
eqs.(\ref{rad_1}, \ref{rad_2}) for $J > 0$. In fact, it follows from
eq.(\ref{singular_lin}) and eq.(\ref{mu_nu}) that the leading singularities
of $\olin{v}_J (p, q)$ and $\olin{\olin{v}}_J (p, q)$, as defined in
eq.(\ref{vector_matel}), are equal to that of $v_J (p, q)$, and that the
leading singularities cancel in $\wtil{v}_J (p, q)$.
\par
For a pure oscillator potential the entire Mult\-hopp integral,
eq.(\ref{multhopp_integral}), can be evaluated analytically. In this case
the Multhopp method provides an alternative to solving the differential
equation for the radial wave functions state by state \cite{leyaouanc_85},
as the solution of eq.(\ref{multhopp}) immediately gives the ground state
and the excited state spectrum.
\par
The scale parameter $\lambda$ is chosen to be of the order of the r.m.s.\
3--momentum of the bound state wave function. Eigenvalues and
eigenfunctions should not be sensitive to this choice. We emphasize that
this method is not a variational approach, \ie , no variation is performed
with respect to $\lambda$, in contrast to the procedure of
\cite{resag_93,lagae_92}.
\section{Normalization and decay constants}
\setcounter{equation}{0}
In this appendix we evaluate the matrix elements needed for the
normalization of the bound state amplitude and for the calculation of the
pseudoscalar meson decay constants within the bilocal field approach. We
shall derive explicit expressions in terms of the bound state wave
functions in the rest frame.
\par
For the normalization of the bound state amplitude we consider the matrix
element of the free part of the quark loop, eq.(\ref{W_2}), between
on--shell bound states $H\sim (q_i \bar{q}_j )$,
\be
\langle H (\cP_H' ) | W_{\rm eff}^{(2)} | H (\cP_H ) \rangle
&=& (2\pi )^4 \delta (\cP_H' - \cP_H ) (2\omega_H' 2\omega_H )^{-1/2}
{\cal G}^{-1} (\cP_H ) , \\[.5ex]
{\cal G}^{-1} (\cP_H ) &=& - i \half N_{\rm c}
\Tr [ \olin{\Gamma} (\cP_H ) G_1 \Gamma (\cP_H ) G_2 ] .
\ee
The trace implies integration over the loop momentum. We evaluate
${\cal G}^{-1}$ in the rest frame of the bound state,
$\cP_H = (M_H, \vec{0})$. Writing
$\Gamma (\bfq | \cP_H ) \equiv  \Gamma (\bfq), \,
\olin{\Gamma} (\bfq | \cP_H )
= \Gamma (\bfq | -\cP_H ) \equiv \olin{\Gamma} (\bfq)$, we have
\be
\lefteqn{ {\cal G}^{-1}(M_H, \vec{0} )
\; = \; - i \half N_{\rm c} \int \frac{d^{4}q}{(2 \pi)^{4}}
\tr\biggl[ \Gamma (\bfq ) G_i ( q - \half\cP_H )
\olin{\Gamma} (-\bfq ) G_j ( q + \half\cP_H ) \biggr]  }
\label{prop_rest_frame} \\
&& = \; - \half N_{\rm c} \mint{q} \tr \biggl[
\frac{\olin{\Gamma}(-\bfq ) \Lambda_{i+} (\gamma_0 \Gamma (\bfq )
\gamma_0 ) \olin{\Lambda}_{j-}}{E_q - M_H} +
\frac{\olin{\Gamma}(-\bfq ) \Lambda_{i-} (\gamma_0 \Gamma (\bfq )
\gamma_0 ) \olin{\Lambda}_{j+}}{E_q + M_H} \biggr] .
\nonumber
\ee
The numerators here can be rewritten as
\be
\lefteqn{
\tr [ \olin{\Gamma}(-\bfq ) \Lambda_\pm^i (\gamma_0 \Gamma (\bfq )\gamma_0 )
\olin{\Lambda}_\mp^j ] \; =}
\nonumber \\
&& - \tr [ (S_j^{-1}(\bfq )\olin{\Gamma} (-\bfq ) S_i^{-1}(\bfq ))
\Lambdaz_\pm (S_i^{-1}(\bfq ) \Gamma (\bfq ) S_j^{-1}(\bfq )) \Lambdaz_\mp ],
\ee
where we have used the relation between the free and the rotated
projectors, eq.(\ref{projectors}). We now use the Salpeter equation, in the
rest frame, eq.(\ref{schroedinger_eq}), and the corresponding equation for
$\olin{\Gamma}(\bfq)$ to express eq.(\ref{prop_rest_frame}) in terms of the
``undressed'' wave function in the rest frame, eq.(\ref{Psi_undressed}),
\be
{\cal G}^{-1}(M_H, \vec{0}) &=& \half N_{\rm c} \mint{q} \Bigl\{
(E_q - M_H) \tr [\Lambdaz_- \olin{\Psiz}(-\bfq ) \Lambdaz_+ \Psi (\bfq )]
\label{prop_Psi} \\
&& + (E_q + M_H) \tr [\Lambdaz_+ \olin{\Psiz}(-\bfq )
\Lambdaz_- \Psi (\bfq )] \Bigr\} .
\nonumber
\ee
In a general frame, $M_H$ is to be replaced by $\sqrt{\cP_H^2}$, and the
integration in eq.(\ref{prop_Psi}) goes over the transverse momentum,
$q^\perp$. The bound state amplitudes are normalized by the requirement
that the dispersion relation be the same as that of a free relativistic
particle,
\be
\cP_\mu \frac{\partial}{\partial\cP_\mu} {\cal G}^{-1}(\cP )|_{\cP^2 = M_H^2}
&=& M_H^2 .
\ee
Note that for a co-moving instantaneous interaction, eq.(\ref{K_eta}), the
interaction part of the effective action, eq.(\ref{W_eff}), does not
contribute to the normalization matrix element because $K^\eta$ depends on
$\cP$ only through the unit boost vector, $\eta$, and $\cP_\mu (\partial
/\partial \cP_\mu) \eta = 0$. From eq.(\ref{prop_Psi}), the normalization
condition in the rest frame can thus be expressed as
\be
1 &=& \frac{N_{\rm c}}{2M_H} \mint{p} \Bigl\{
\tr [\Lambdaz_- \olin{\Psiz}(-\bfq ) \Lambdaz_+ \Psiz (\bfq )]
- \tr [\Lambdaz_+ \olin{\Psiz}(-\bfq ) \Lambdaz_- \Psiz (\bfq )] .
\Bigr\}
\ee
In terms of the components of the wave function, eq.(\ref{Psi_L_N}), this
reads
\be
1 &=& \frac{2 N_{\rm c}}{M_H} \mint{p} \left(
L^{(1)*} L^{(2)} + L^{(2)*} L^{(1)}
+ \bfN^{(1)*} \!\cdot\! \bfN^{(2)} +
\bfN^{(2)*} \!\cdot\! \bfN^{(2)} \right) ,
\label{L_N_normalization}
\ee
where $L^{(1)*} \equiv L^{(1)*}(\bfq ) = L^{(1)}(-\bfq )$, {\em etc.}.
\par
We now consider the pseudoscalar meson decay constants. The matrix element
of eq.(\ref{W_semi_2}) for the decay of a meson $H \sim (q_i \bar{q}_j )$
into a leptonic pair is given by
\be
\lefteqn{
\matel{l \nu (\cP_L )}{W^{(2)}_{\rm semi}}{H(\cP_H )}
\; =\; (2 \pi )^4 \delta^{(4)}(\cP_{H} -\cP_{L} ) \frac{G_{F}}{\sqrt{2}}
\matel{l \nu}{l_\mu}{0} {\cal F}^\mu (\cP_H ) ,}
\label{meson_decay_matel} \\
&& {\cal F}^\mu (\cP_H ) \; =\; -i N_{\rm c}
\int\frac{d^{4}q}{(2 \pi)^{4}} \tr_\gamma
[ O^{\mu} G_{i} ( q - \half\cP_{H}) \olin{\Gamma} (q^\perp \vert \cP_{H})
G_{j} (q + \half\cP_{H}) ] . \hspace{1.5em}
\ee
By integrating over the parallel component of the loop momentum, this
matrix element can be evaluated in terms of the wave function of the
``moving'' bound state, eq.(\ref{Psi_dressed_general}),
\be
{\cal F}^\mu (\cP_H ) &=& N_{\rm c} \mint{q^\perp} \tr_\gamma
[ O^\mu \olin{\Psi} (q^\perp \vert \cP_{H}) ] ,
\label{F_Psi}
\ee
where
$\olin{\Psi} (q^\perp \vert \cP_{H}) \equiv \Psi(q^\perp \vert -\cP_{H})$.
The decay constant is then read off by comparing eq.(\ref{meson_decay_matel})
with the general definition
\be
\matel{l \nu (\cP_L ) }{W^{(2)}_{\rm semi}}{H_{ij} (\cP_H )} &=&
 (2 \pi )^{4} \delta^{4}(\cP_H -\cP_L ) \frac{G_{F}}{\sqrt{2}} F_{ij}
\cP^{\mu}_{H} \matel{l \nu}{l_{\mu}}{0} .
\ee
Note that for 3--dimensional interactions of the form eq.(\ref{K_eta}) in
general different values for the timelike and spacelike part of the pion
decay constant are obtained \cite{leyaouanc_85}.
\par
In this paper we only consider the decays of pseudoscalar mesons ($J^P =
0^-$) at rest, $\cP_H = (M_H, \vec{0})$. In this case the wave function has
the form
\be
\olin{\Psiz} ({\bf q}) &=& \left\{ L^{(1)} ({\bf q})
- \gamma_0 L^{(2)} ({\bf q}) \right\} \gamma^{5} ,
\ee
as can be seen from the partial--wave expansion in appendix A. Inserting
this expression with eq.(\ref{Psi_undressed}) into eq.(\ref{F_Psi}) and
calculating the trace we obtain the decay constant of a pseudoscalar meson
$H \sim (q_i \bar{q}_j )$ at rest,
\be
F_H &=& \frac{4N_{c}}{M_H} \mint{q}
L^{(2)} ({\bf q}) \cos (\nu_{iq} + \nu_{jq}) .
\label{decay_constant}
\ee
Here, the wave functions in the rest frame $L^{(1, 2)}(\bfq )$ satisfy
eqs.(\ref{L_eq}, \ref{N_eq}) and are normalized according to
eq.(\ref{L_N_normalization}). In particular, in the isospin limit,
$\nu_{iq} = \nu_{jq} = \half (\half\pi - \phi (\magq ) )$, the decay
constant becomes
\be
F_H &=& \frac{4N_{c}}{M_H} \mint{q} L^{(2)} ({\bf q}) \sin \phi (\magq ) .
\label{fpi}
\ee
For light mesons ($\pi , K$) it is customary to define the decay constant as
$f_\pi = F_\pi /\sqrt{2}$, {\em etc.}.
%
%

%
\newpage
\noindent {\Large\bf Tables}
\\[.4cm]
%
%
\begin{table}[h]
\centering
\begin{tabular}{|c|c|c|c|c|}
\hline
 & dynamical & $m_q = 0.082 $ & $m_q = 0.33$
& exp. \\ \hline
\hline
$m_D$    &  2.03  &  2.33  &  2.34  & 1.87     \\ \hline
$F_D$    &  0.16  &  0.21  &  0.25  &  $\leq$ 0.29 \\ \hline
$m_{D'}$ &  2.54  &  2.85  &  2.89  &          \\ \hline
$F_{D'}$ &  0.16  &  0.16  &  0.19  &          \\ \hline
\hline
$m_B$    &  5.38  &  5.64  &  5.65  & 5.28     \\ \hline
$F_B$    &  0.18  &  0.24  &  0.29  &          \\ \hline
$m_{B'}$ &  5.83  &  6.11  &  6.14  &          \\ \hline
$F_{B'}$ &  0.21  &  0.23  &  0.26  &          \\ \hline
\hline
\end{tabular}
\caption[]{The masses and decay constants of the $0^-$ $D$-- and $B$--mesons
and their first radially excited states for a linear plus Coulomb potential
with $\sigma = 0.41\,{\rm GeV}, \, \alpha_s = 0.39, \, m_c = 1.39\,{\rm
GeV}$ and $m_b = 4.79\,{\rm GeV}$. All energies in GeV. Masses and wave
functions were determined from eq.(\ref{L_eq}), decay constants from
eq.(\ref{decay_constant}). Columns 1--3 refer to different descriptions of
the light quark spectrum. 1: dynamical chiral symmetry breaking,
eq.(\ref{sde_system}), with $m^0 = 1\,{\rm MeV}$, 2: constant light quark
mass, eq.(\ref{constant_mass}), $m_q = m(\magq = 0) = 0.082\,{\rm GeV}$, 3:
constant light quark mass, $m_q = 0.33\,{\rm GeV}$. The experimental bound
on $F_D$ in column 4 is from ref.\ \cite{markIII_88}.}
\label{tab_data}
\end{table}
%
%
\begin{table}[h]
\[
\renewcommand{\arraystretch}{2.}
\begin{array}{|c|c|c|}
\hline
V(r) & \Vpq & v_L (p, q)
\\ \hline
1/r & {\displaystyle \frac{4\pi}{|\bfp - \bfq |^2} } &
{\displaystyle \frac{1}{\pi} Q_L (x) }\\
r & {\displaystyle - \frac{8\pi}{|\bfp - \bfq|^4}
+ C \delta^{(3)}(\bfp - \bfq ) } &
{\displaystyle \frac{1}{\pi}\frac{1}{pq} \frac{d Q_L}{dx}(x) }
\\[1.5ex]
r^2 & {\displaystyle (2\pi )^3 \nabla^2_\bfq \delta^{(3)}(\bfp - \bfq ) } &
{\displaystyle
\left( \frac{d^2}{dq^2} + \frac{L(L + 1)}{q^2} \right) \delta (p - q) }
\\[1.5ex]
\hline
\end{array}
\renewcommand{\arraystretch}{1.}
\]
\caption[]{The Fourier transforms and the angular matrix elements,
eq.(\ref{v_l}), corresponding to the power--like potentials used in
eq.(\ref{potential}). The linear potential involves a contact term, $C =
8\pi\mint{q} \magq^{-4}$, which cancels the IR--divergence. Here, $Q_L (x)$
is the Legendre function of the second kind, with $x = (p^2 + q^2)/2pq$.}
\label{tab_potential}
\end{table}
%
%
\newpage
\noindent {\Large\bf Figure captions}
\\[.4cm]
Fig.1: The chiral angle, $\phi (\magp )$, from the Schwinger--Dyson
equation for a light flavor ($m^0$ = 1 MeV). Solid line: linear plus
Coulomb potential ($\sigma = 0.41\,{\rm GeV},\, \alpha_s = 0.39$), dotted
line: pure oscillator potential ($V_0 = 0.247\,{\rm GeV}$). The dashed line
shows the value for a constant heavy quark mass, $m_Q = m_c = 1.39\, {\rm
GeV}$.
\\[.6cm]
Fig.2: The quark energy, $E(\magp )$, from the Schwinger--Dyson equation
for a light flavor ($m^0$ = 1 MeV). The potentials are those of
fig.1. The dashed line shows the energy for a constant heavy quark
mass, $m_Q = m_c = 1.39\, {\rm GeV}$. The straight solid line indicates the
asymptotic behavior, $E = \magp$.
\\[.6cm]
Fig.3: The dynamical quark mass, $m(\magp )$ of eq.(\ref{dynamical_mass}),
for a light flavor ($m^0$ = 1 MeV). The potentials are those of fig.1.
\\[.6cm]
Fig.4: The squared pion mass, $m_\pi^2$, in the chiral limit, as a function
of the current quark mass, $m^0_u = m^0_d \equiv m^0$. The potentials are
to those of fig.1. Solid line: linear plus Coulomb potential, dotted line:
oscillator potential.
\\[.6cm]
Fig.5: The decay constants of the pion, $f_\pi$, and its first radially
excited state, $f_{\pi'}$, in the chiral limit. The potentials are those of
fig.1. Solid lines: linear plus Coulomb potential, dotted lines: oscillator
potential. Note the different scales for $f_\pi$ and $f_{\pi'}$. Here,
$f_\pi = F_\pi / \sqrt{2}, f_{\pi'} = F_{\pi'} / \sqrt{2}.$
\\[.6cm]
Fig.6: The radial wave functions of the pion, $\ell^{(1, 2)}(\magp )$, for
a linear plus Coulomb potential ($\sigma = 0.41\, {\rm GeV},\, \alpha_s =
0.39$) and $m^0_u = m^0_d = 1\,{\rm MeV}$. Solid lines: $\ell^{(1)}$,
dotted lines: $\ell^{(2)}$. Shown are the ground and first excited state.
The single-particle spectrum is determined by the Schwinger--Dyson
equation.
\\[.6cm]
Fig.7: The (1)--component of the radial wave function of the $D$--meson,
$\ell^{(1)}(\magp )$, for a linear plus Coulomb potential ($\sigma =
0.41\,{\rm GeV},\, \alpha_s = 0.39$) and $m_c = 1.39\,{\rm GeV}$. Solid
line: dynamical quark self--energy for the light flavor ($m^0 = 1\, {\rm
MeV}$), dotted line: constant light quark mass, $m_q = 0.082\,{\rm GeV}$.
We have not shown $\ell^{(2)}$; it is very close to $\ell^{(1)}$ due to the
large mass of the charmed quark.
\end{document}